\documentclass{ar-1col-S2O}
\usepackage{url}
\usepackage{bm}
\usepackage[numbers]{natbib}
\usepackage{amsmath}
\usepackage[normalem]{ulem}
\usepackage{comment}

\jname{Xxxx. Xxx. Xxx. Xxx.}
\jvol{AA}
\jyear{YYYY}
\doi{((please add article doi))}

\begin{document}


\markboth{Author et al.}{Short title}

\title{Neutrino Oscillations in Core-Collapse Supernovae and Neutron Star Mergers}

\author{Lucas Johns,$^1$ Sherwood Richers,$^2$ and Meng-Ru Wu$^{3,4,5}$ 
\affil{$^1$Theoretical Division, Los Alamos National Laboratory, Los Alamos, NM 87545, USA; email: ljohns@lanl.gov}
\affil{$^2$Department of Physics and Astronomy, University of Tennessee, Knoxville, USA; email: richers@utk.edu}
\affil{$^3$Institute of Physics, Academia Sinica, Taipei 115201, Taiwan; email: mwu@gate.sinica.edu.tw}
\affil{$^4$Institute of Astronomy and Astrophysics, Academia Sinica, Taipei 106319, Taiwan}
\affil{$^5$Physics Division, National Center for Theoretical Sciences, Taipei 106319, Taiwan}}

\begin{abstract}
Accurate neutrino transport is crucial for reliably modeling explosive astrophysical events like core-collapse supernovae (CCSNe) and neutron star mergers (NSMs). However, in these extremely neutrino-dense systems, flavor oscillations exhibit challenging nonlinear effects rooted in neutrino--neutrino forward scattering. Evidence is quickly accumulating that these collective phenomena can substantially affect explosion dynamics, neutrino and gravitational-wave signals, nucleosynthesis, and kilonova light curves. We review the progress made so far on the difficult and conceptually deep question of how to correctly include this physics in simulations of CCSNe and NSMs. Our aim is to take a broad view of where the problem stands, and so provide a critical assessment of where it is headed.
\end{abstract}

\maketitle

\tableofcontents

\section{INTRODUCTION}

During the decades-long saga of the solar neutrino problem, neutrino flavor oscillations went from speculative hypothesis to uncontested fact. The present-day quantitative agreement between predictions and observations of solar neutrinos rests on the theory of neutrino propagation. Neutrinos detected from the Sun corroborate our understanding of how the solar medium influences neutrino flavor conversion.

Neutrino transport theory becomes much less settled when we shift our attention to the more extreme stellar environments of core-collapse supernovae (CCSNe) and neutron star mergers (NSMs). These are two types of events of great importance in the era of multimessenger astronomy, but we do not yet know how to correctly incorporate oscillations into the relevant predictions. The growing evidence that oscillations meaningfully alter the modeling of these sites has moved this open question---what we call the \textbf{oscillation problem}---to the forefront of theoretical astrophysics.

The challenge comes from the nonlinearity of oscillations in highly neutrino-dense settings. Forward scattering of neutrinos on background matter, which modifies neutrino propagation in the Sun, gives rise to a relatively simple phenomenology. Forward scattering of neutrinos \textit{on each other}, with its array of collective effects, adds a great deal more depth and complexity to the flavor dynamics. Significant progress has been made on this topic over the past several years. Various strategies for solving the oscillation problem---for including oscillations into predictions of explosion dynamics, neutrino signals, nucleosynthesis yields, and kilonova light curves---are now being devised and implemented.

The focus of this review is the intense ongoing effort to solve the oscillation problem. Another review in the same volume of this journal covers other aspects of supernova modeling \cite{janka2025long}. Our objective is to provide a useful guide, not an exhaustive one. Classic reviews from 2010 \cite{duan2010collective} and 2016 \cite{mirizzi2016supernova} remain valuable resources on collective neutrino oscillations. A recent comprehensive review \cite{volpe2024neutrinos} touches on a number of topics largely omitted here, and several others particularly offer greater coverage of nucleosynthesis \cite{fischer2024neutrinos}, flavor instabilities \cite{chakraborty2016collective, tamborra2021new, richers2023fast}, many-body correlations \cite{patwardhan2022many}, the early universe \cite{capozzi2022neutrino}, and neutrino astronomy \cite{tamborra2024neutrinos}.

We begin with an exposition of the motivations and challenges from the standpoint of astrophysics (Sec.~\ref{sec:astro}). We next discuss the formalism and foundations of neutrino quantum kinetics (Sec.~\ref{sec:fields}) and present the major concepts and themes that have emerged from studies of neutrino flavor evolution (Sec.~\ref{sec:essential}). We then summarize the approaches being developed for integrating oscillations into astrophysical simulations (Sec.~\ref{sec:strategies}).

\section{ASTROPHYSICAL MOTIVATIONS\label{sec:astro}}

\subsection{Mesoscopic dynamics in macroscopic settings}

The gravitational collapse of a massive star compresses matter to extreme densities and temperatures. As the collapsed stellar core neutronizes and cools, it emits a tremendous flux of neutrinos. These particles strongly influence the chemical composition and in most cases drive the explosion. As SN1987A demonstrated, they are also detectable.

In the merger of two neutron stars, matter begins at extreme densities and is then violently heated and reconfigured. The formation and relaxation of the post-merger compact object and surrounding accretion disk often entail prodigious neutrino emission. The effect on nucleosynthesis can be dramatic. Moreover, the nucleosynthetic products are observable through their radioactive decay, as exemplified by AT2017gfo, the kilonova that accompanied GW170817.

\begin{figure}
    \centering
    \includegraphics[width=0.60\linewidth]{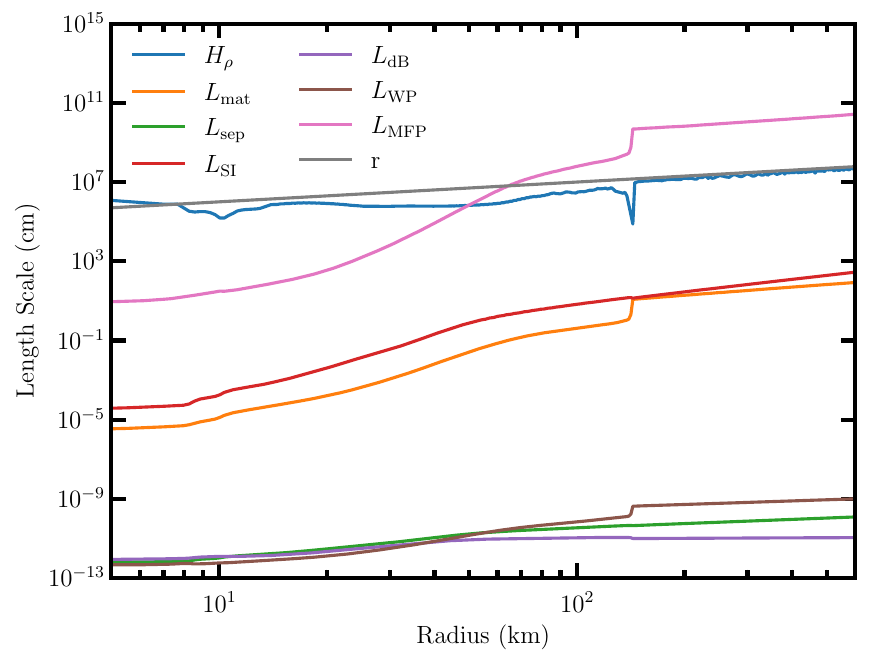}
    \caption{Length scales in a 1D CCSN simulation at the time of maximum shock radius. $H_\rho=\rho (d\rho/dr)^{-1}$ is the density scale height. $L_\mathrm{mat}=(\sqrt{2} G_F n_e)^{-1}$ is the oscillation length scale associated with forward scattering on background electrons. $L_\mathrm{sep}=n_{\nu_e}^{-1/3}$ is the interparticle separation. $L_\mathrm{SI}=(\sqrt{2} G_F n_{\nu_e})^{-1}$ is the length scale associated with forward scattering on background neutrinos. $L_\mathrm{dB}=2\pi (\langle E_{\nu_e} \rangle)^{-1}$ is the neutrino de Broglie wavelength. $L_\mathrm{WP}$ is the estimated size of a neutrino wave packet produced in that region \cite{kersten2016decoherence, akhmedov2017collective}. $L_\mathrm{MFP}=\langle \kappa_\mathrm{abs}+\kappa_\mathrm{scat}\rangle^{-1}$ is the $\nu_e$ mean free path. $r$ is the radial coordinate.}
    \label{fig:length_scales}
\end{figure}

Radiation-hydrodynamic simulations are the vanguard of CCSN/NSM theory. They are largely concerned with the \textbf{macroscopic scale} (Fig.~\ref{fig:length_scales}). CCSNe and NSMs host compact cores that are $\mathcal{O}(10)$~km in radius. In the case of a merger, the accretion disk may extend out $\mathcal{O}(100)$~km. In a CCSN, during the decisive first few hundred milliseconds after core bounce, the radius of the shock wave is $\mathcal{O}(100)$~km. In the dense interiors of these sites, all particles with the exception of the weakly interacting neutrinos are treated using hydrodynamics. Where small-scale features do occur in the fluid dynamics---in turbulent regions and the shock---they are not fully resolved. Neutrinos, which are not in thermal equilibrium with the fluid, are treated using kinetics or some approximation of it. During the transition from trapped to free-streaming, their mean free path is $\mathcal{O}(10)$~km, hence also macroscopic.

Neutrino transport is computationally demanding even without oscillations. Evolving phase-space distribution functions under Boltzmann collision integrals carries a high cost. Oscillations make the numerical challenges far steeper by introducing a \textbf{mesoscopic scale} represented by the oscillation length. For typical neutrino energies, the oscillation length in vacuum is at the kilometer scale and upward. But when neutrinos are immersed in a dense bath of electrons and other particles, their oscillations are modified by potentials arising from forward scattering \cite{wolfenstein1978neutrino, mikheyev1985resonance, notzold1988neutrino, pantaleone1992neutrino} (Sec.~\ref{sec:processes}). In the densest regions, the oscillation length in the medium is many orders of magnitude smaller than in vacuum. Running a simulation with this level of resolution is plainly untenable. The oscillation problem, in short, is to figure out how to accommodate in-medium neutrino flavor mixing without fully resolving it. In this respect, it resembles the problem of modeling (magneto)hydrodynamic turbulence.

The \textbf{microscopic scale} is down many orders of magnitude further, well below the oscillation length. Here we find the interparticle separation, the de Broglie wavelength, and the size of a neutrino wave packet. Neutrino quantum kinetics is a mesoscopic theory. The microscopic scale becomes relevant when considering the foundations of quantum kinetics in quantum many-body theory (Sec.~\ref{sec:fields}).

\subsection{Prevalence of flavor instabilities\label{sec:prevalence}}

Neutrino oscillations introduce mesoscale dynamics. Linear stability analysis \cite{banerjee2011linearized, izaguirre2017fast} allows us to begin addressing the obvious follow-up question: How important are oscillations? Flavor instabilities that appear in post-processed simulation data are evidence that oscillations, if they had been included in the simulation, would have caused some amount of deviation from the results obtained without them. A complete picture of the prevalence of instabilities is still being developed, but already we can see that they are widespread.

\begin{figure}
    \centering
    \begin{minipage}{0.46\linewidth}
    \includegraphics[width=\linewidth]{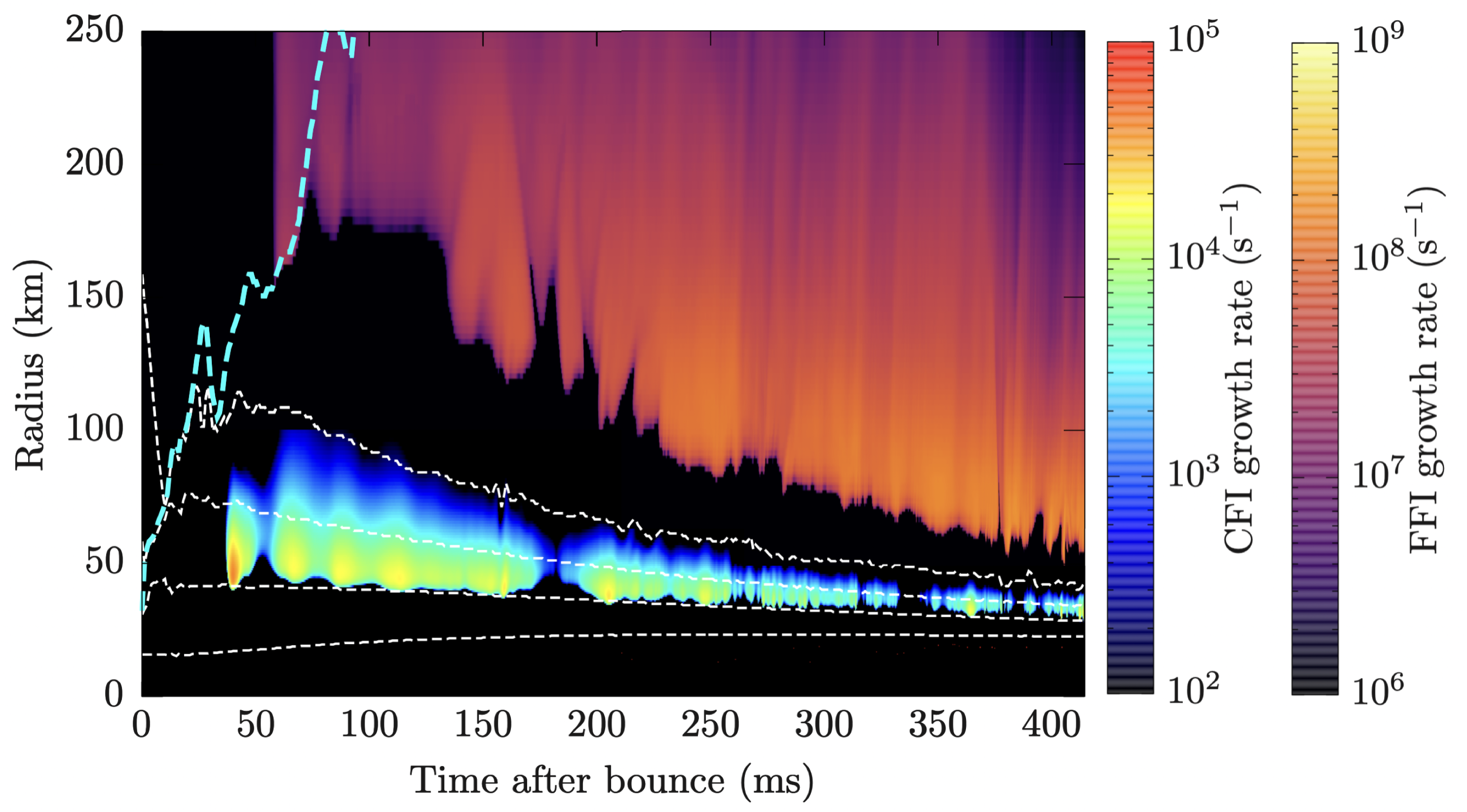}
    \end{minipage}
    \begin{minipage}{0.50\linewidth}
    \includegraphics[width=\linewidth]{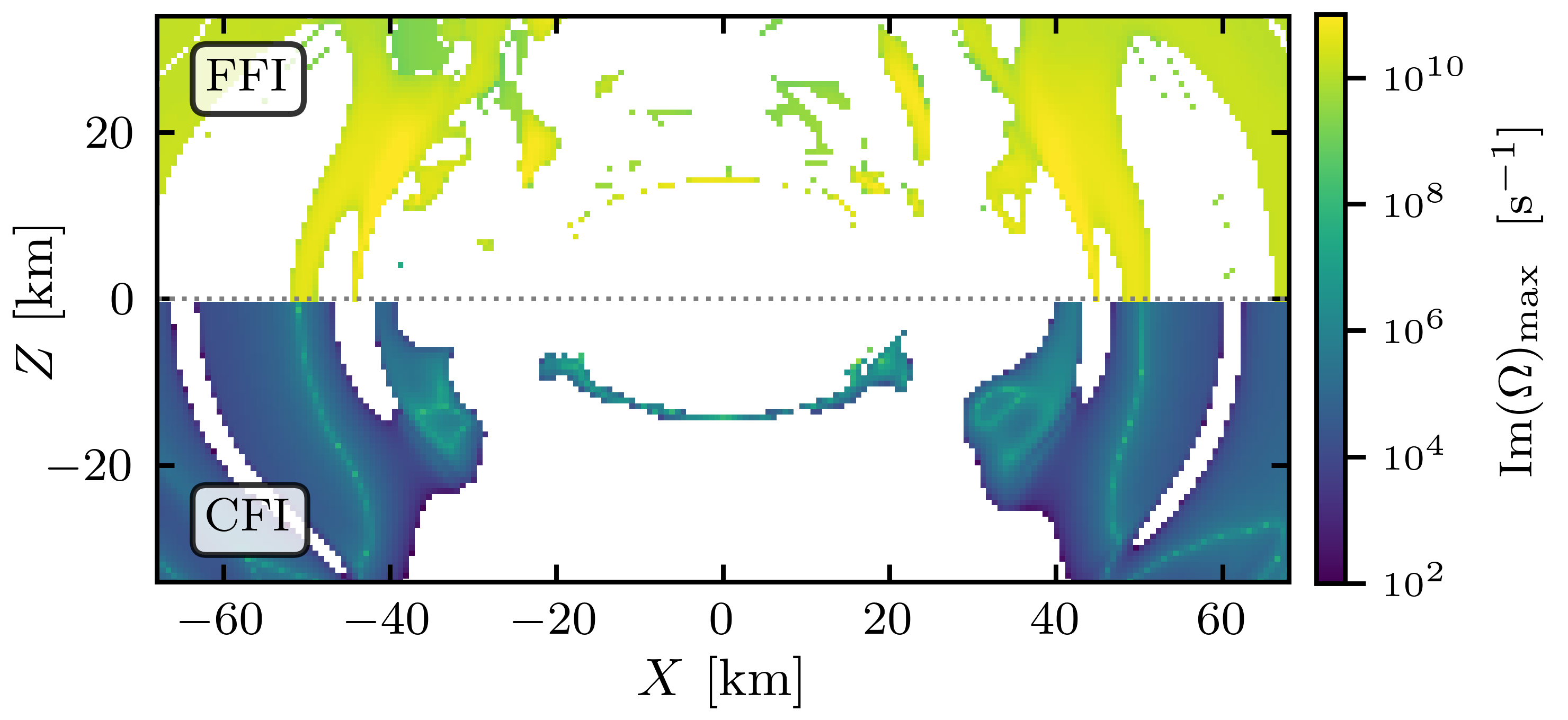}
    \end{minipage}
    \caption{\textit{Left}: Growth rates of FFI and CFI in a 2D CCSN simulation. The dashed white lines are density contours of $10^{10}$, $10^{11}$, $10^{12}$, and $10^{13}$~g~cm$^{-3}$. The dashed cyan curve shows the location of the shock. Adapted with permission from Akaho et al. (2024) \cite{akaho2023collisional}. \textit{Right:} Growth rates of FFI and CFI in a simulation of two merging neutron stars. The slice shown is in the equatorial plane, 5~ms after merger. Alternating zones of stability/instability follow spiral density waves in the disk. Adapted with permission from Froustey et al. (2024) \cite{froustey2024neutrino}.}
    \label{fig:prevalence}
\end{figure}

A turning point in this research area was the realization that the conditions for \textbf{fast flavor instability (FFI)} \cite{sawyer2009multiangle, sawyer2016neutrino} are met over a range of times and locations relevant for CCSN and NSM observables \cite{wu2017fast, abbar2019occurrence, delfan2019linear, morinaga2020fast, abbar2021characteristics, nagakura2021where, li2021neutrino, fernandez2022fast} (Fig.~\ref{fig:prevalence}). FFI occurs when the angular distributions of $\nu_e$ and $\bar{\nu}_e$ are sufficiently distinct. The angular distributions of all species are nearly isotropic in the neutrino trapping region and become increasingly forward-peaked as a function of distance from the principal emitting regions (the compact object and/or accretion disk). There is a species-dependence overlaid on this trend that has two principal origins. Firstly, charged-current processes involving nucleons generate imbalances that reflect the neutron-to-proton ratio. For example, in neutron-rich environments, neutrino absorption on nucleons delays the decoupling of $\nu_e$ relative to $\bar{\nu}_e$. Secondly, neutral-current scattering acts more strongly on $\bar{\nu}_e$ than $\nu_e$ due to the former's higher average energy. These factors tend to deplete $\nu_e$ in outgoing directions and enhance $\bar{\nu}_e$ in ingoing directions. The result is that FFI is pervasive in multidimensional simulations, with an associated length scale commonly on the order of centimeters.

Evidence indicates that \textbf{collisional flavor instability (CFI)} \cite{johns2021collisional} is also prevalent \cite{xiong2022evolution,xiong2023collisional,liu2023universality, akaho2023collisional, shalgar2023neutrinos, froustey2024neutrino}. This type of instability results from sufficiently distinct collision rates of $\nu_e$ and $\bar{\nu}_e$. The prime locations for CFI are the neutrino decoupling regions. In trapping regions, flavor instabilities of all kinds tend to be suppressed by the electron-flavor chemical potential $\mu_{\nu_e}$. In free-streaming regions, collisional rates are so low that CFI is probably unimportant even where it does appear. But in between, CFI is common and typically grows on a length scale of $\mathcal{O}(10)$~km. This scale can be much smaller when CFI is resonant \cite{xiong2023collisional, liu2023systematic}.

Although \textbf{slow flavor instability (SFI)} \cite{kostelecky1993neutrino, duan2006collective} was the first type to be discovered, we know the least about its prevalence. Recent work has stressed the need for meticulous studies of the kinds that have been done for FFI \cite{neto2023energy, shalgar2024neutrino, fiorillo2024theoryslow, fiorillo2025theoryslow}.

In fact, even analyses of CFI have mostly been based on the approximation of isotropic angular distributions and a restriction to the homogeneous (wave vector $\bm{K} = 0$) collective mode. A significant objective over the coming years will be to bring our knowledge of CFI and SFI up to par with FFI. At the same time, further progress can be made on the prevalence of FFI as well. Analyses of simulations with high angular resolution, or analyses where angular-moment data are handled carefully, are especially valuable, as findings are compromised by low angular resolution \cite{johns2021fast, nagakura2021constructing, mukhopadhyay2024time}. For all instabilities, more work on three-flavor mixing is desirable \cite{capozzi2020mu, capozzi2021fast, shalgar2021three, liu2024muon}.

Based on linear stability analysis, oscillations are expected to cause deviations from the results produced by state-of-the-art simulations. We now turn to the question of what the effects might be.

\subsection{Estimated effects of oscillations\label{sec:effects}}

\begin{figure}
    \centering
    \begin{minipage}{0.32\linewidth}
    \includegraphics[width=\linewidth]{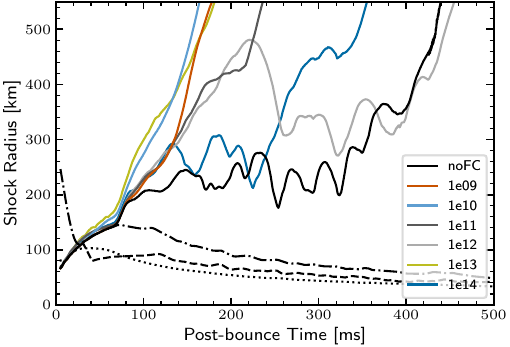}
    \end{minipage}
    \begin{minipage}{0.35\linewidth}
    \includegraphics[width=\linewidth]{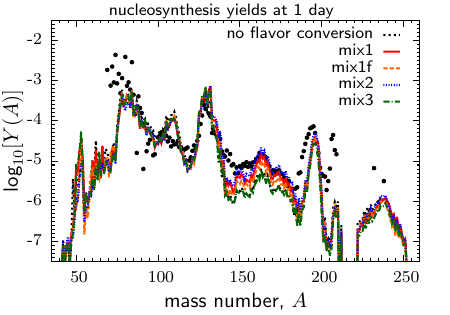}
    \end{minipage}
    \begin{minipage}{0.29\linewidth}
    \includegraphics[width=\linewidth]{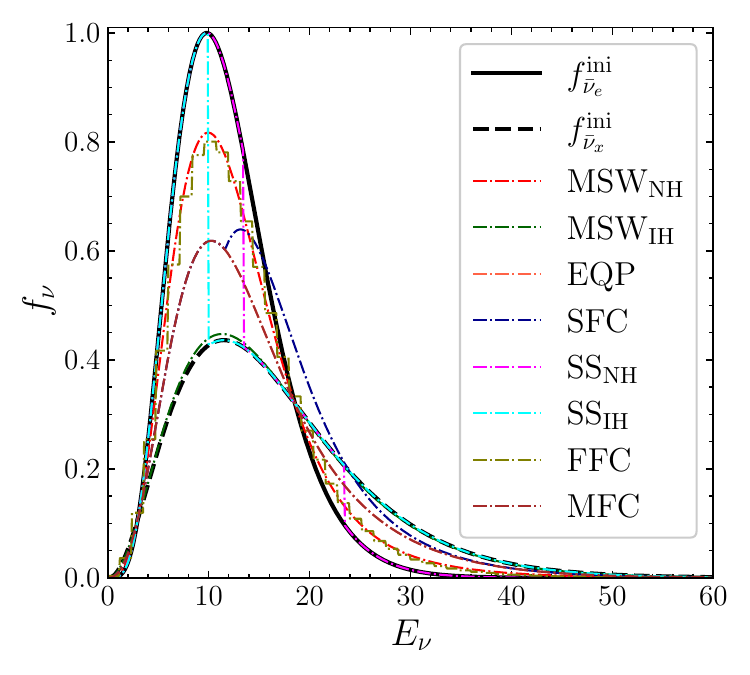}
    \end{minipage}
    \caption{Potential astrophysical effects of neutrino oscillations. \textit{Left:} Angle-averaged shock radius in 2D CCSN simulations with flavor conversion setting in above different critical densities (color-coded, with labels in g~cm$^{-3}$), showing significant deviations from the simulation without oscillations (``noFC''). Adapted with permission from Ehring et al. (2023) \cite{ehring2023fastneutrino}. \textit{Middle:} Nucleosynthesis abundance distributions formed in the simulated ejecta of a post-merger accretion disk, for various flavor-mixing prescriptions. Enhanced lanthanide abundances are observed. Modified with permission from Just et al. (2022) \cite{just2022fast}. \textit{Right:} CCSN neutrino energy spectra in various possible flavor-mixing scenarios, with potentially observable consequences. Adapted with permission from Abbar and Volpe (2024) \cite{abbar2024using}.}
    \label{fig:astro_effects}
\end{figure}

If there were no influence from collective oscillations, in-medium flavor conversion would proceed solely through the \textbf{MSW effect} \cite{mikheyev1985resonance}. The underlying mechanism is adiabatic tracking of energy eigenstates, which vary as a function of electron density $n_e$ (Sec.~\ref{sec:adiabaticity}). MSW conversion is well understood, with subtleties arising only in cases where the gentle variation of $n_e$ is disrupted in the resonance regions by a shock \cite{schirato2002connection} or turbulent fluctuations \cite{duan2009neutrino}. Except perhaps in low-mass CCSNe \cite{duan2008flavor}, collective oscillations are indeed stifled during the neutronization burst by the stark hierarchy of neutrino number densities $n_{\nu_e} \gg n_{\nu_\mu} \gg n_{\bar{\nu}_e}$. The MSW effect is the lone signature of oscillations during this stage. After the first several tens of milliseconds, however, flavor conversion of uncertain outcome occurs well before neutrinos reach the resonance regions.

Not all collective phenomena are instabilities. \textbf{Spectral swaps} \cite{duan2006simulation, duan2006coherent} and \textbf{matter--neutrino resonant conversion} \cite{malkus2012neutrino, malkus2014matter} are, like the MSW effect, consequences of adiabaticity (Sec.~\ref{sec:adiabaticity}). Being predictions of highly simplified astrophysical models, their status in real CCSNe and NSMs is currently unclear. Additionally, and again like the MSW effect, they are typically downstream of flavor instabilities. Nonetheless, estimates have been made of their possible influence on explosion, nucleosynthesis, and neutrino signals \cite{duan2011influence, choubey2010signatures, dasgupta2012role, wu2015effects, sasaki2017possible, ko2020neutrino, stapleford2020coupling}.

Flavor instabilities open the door to much more substantial effects on the astrophysics. For instance, as we saw in Sec.~\ref{sec:prevalence}, flavor conversion likely occurs behind the supernova shock. Estimates indicate that \textbf{explosion dynamics} could be significantly affected \cite{ehring2023fast, ehring2023fastneutrino, nagakura2023roles, mori2025three, wang2025effect}. The left panel of Fig.~\ref{fig:astro_effects} exemplifies the findings obtained using approximate treatments of oscillations in CCSN simulations. Notably, the evolution of the shock is highly variable depending on precisely where the flavor conversion sets in. This observation suggests that rough approximations, though clearly valuable, are not an adequate answer to the oscillation problem. The details matter.

Flavor instabilities might augur significant changes to \textbf{nucleosynthesis} \cite{wu2017imprints, xiong2020potential, li2021neutrino, just2022fast, fujimoto2023explosive, qiu2025neutrino, lund2025angle} (middle panel of Fig.~\ref{fig:astro_effects}). For example, flavor conversion in a CCSN may substantially affect neutrino-induced nucleosynthesis in outer stellar envelopes~\cite{wu2015effects,ko2020neutrino}. A strong influence is also possible in the outflows from post-merger accretion disks, which are sensitive to neutrinos emanating from the accretion disk and the central neutron star if one exists. In this case, flavor conversion alters the electron fraction in the disk outflow and affects the yield of heavy neutron-rich elements \cite{george2020fast, li2021neutrino, just2022fast,fernandez2022fast, qiu2025neutrino, lund2025angle}. The modification of nuclide abundances could in turn appreciably affect the \textbf{kilonova light curve} produced by the decay of radioactive elements \cite{li2021neutrino, just2022fast}. Firming up the quantitative connection between flavor conversion and kilonovae is a high priority for this research area.

Oscillations are imprinted on the \textbf{neutrino signals} arriving at Earth \cite{gava2009dynamical, abbar2024using} (right panel of Fig.~\ref{fig:astro_effects}). The frequency of CCSNe near enough to be detected in neutrinos is low, but the amount of information that could be extracted from an event in or near our Galaxy is tremendous. A discernible split in the energy spectrum would be a definitive signature of a spectral swap. Whether collective oscillations realistically produce such features is another matter. Various flavor-conversion outcomes are potentially distinguishable, but it is difficult to make broad pronouncements because inferences about oscillations are generally contingent on other knowledge about the event.

The influence of neutrino flavor conversion on \textbf{gravitational-wave emission} is only just beginning to be assessed, but here again there are tantalizing hints that the impacts may be appreciable \cite{ehring2024gravitational, qiu2025neutrino}. In a CCSN, greater emission could result if, for example, protoneutron star convection is enhanced by flavor conversion in the neutrino decoupling region. In a NSM, it could result, for example, from an accelerated contraction of the outer layers of the remnant.

In light of the findings summarized above, solving the oscillation problem has become an urgent task for CCSN/NSM theory. The following sections review the progress made toward this end.

\section{FROM QUANTUM FIELDS TO QUANTUM KINETICS\label{sec:fields}}

\subsection{Quantum coherence and Wigner functions\label{sec:coherence}}

Neutrino transport is fundamentally rooted in quantum field theory. For each mass state $i$, a Dirac neutrino field can be expanded as
\begin{equation}
    \nu_i (t,\bm{x}) = \sum_h \int \frac{d^3\bm{q}}{(2\pi)^3} e^{i \bm{q}\cdot\bm{x}} \left( a_{i,h}(t,\bm{q}) u_{i,h}(\bm{q}) + b_{i,h}^\dagger (t,-\bm{q}) v_{i,h} (-\bm{q}) \right),
\end{equation}
where $u$ and $v$ are Dirac spinors, $a$ is the neutrino annihilation operator, $b^\dagger$ is the antineutrino creation operator, and the sums are over helicity $h$ and momentum $\bm{q}$. (For Majorana neutrinos, there is no independent antineutrino field mode $b$.) The evolution of $a$ and $b$ under the Heisenberg equation contains all of the dynamics of $\nu$.

However, solving for the complete quantum evolution is inconceivable from a practical standpoint. The spirit of quantum kinetics is to be selective in retaining and evolving information characterizing the field. Expectation values of the form $\langle a^\dagger (t) a (t) \rangle$ are generalized occupation numbers, with the generalization referring to the fact that $a^\dagger$ and $a$ may differ in their mass states, helicities, or momenta. In such cases, and in the chosen basis, $\langle a^\dagger (t) a (t) \rangle$ is not a density of particles per se but rather a density of \textbf{quantum coherence}. This review is concerned with mass coherence, or nonzero expectation values of the type $\langle a^\dagger_j (t) a_i (t) \rangle$ for $j \neq i$. Helicity coherence, $\langle a^\dagger_{h'} (t) a_{h} (t) \rangle \neq 0$ for $h' \neq h$, and opposite-momentum pairing, $\langle b (t, -\bm{q}) a(t, \bm{q}) \rangle \neq 0$, are generally suppressed by the high scale of neutrino kinetic energy \cite{volpe2013extended, serreau2014neutrino, cirigliano2015new, kartavtsev2015neutrino, fiorillo2024collective}.

Some amount of momentum coherence must be tolerated to allow for spatial inhomogeneity. To this end, the \textbf{Wigner function} (conventionally called the \textit{density matrix}) is introduced \cite{raffelt1993non, sigl1993general, stirner2018liouville}:
\begin{equation}
    \rho_{\bm{q},ij} (t,\bm{x}) = \int \frac{d^3\bm{\delta q}}{(2\pi)^3} e^{i \bm{\delta q}\cdot\bm{x}} \left\langle a_j^\dagger \left(t,\bm{q}-\frac{\bm{\delta q}}{2} \right) a_i \left(t, \bm{q}+\frac{\bm{\delta q}}{2} \right) \right\rangle.
\end{equation}
Assuming a two-point correlator proportional to $\delta^3 (\bm{\delta q})$ would imply $\bm{x}$-independent $\rho$. Instead of making this strict assumption, we permit coherence of unequal-momentum field modes but assume that $\rho$ changes gradually as a function of $\bm{x}$ relative to the de Broglie wavelength. In effect, the slow-variation assumption coarse-grains over (presumably inconsequential) microscopic features while retaining the astrophysically relevant forms of quantum coherence.

\subsection{Coherent and incoherent processes\label{sec:processes}}

The complete equations of motion for the neutrino field are determined by the Standard Model Lagrangian with the addition of neutrino mass. However, approximations are required to isolate the evolution of $\rho$ from higher-point (multi-particle) correlation functions like $\langle a^\dagger a^\dagger a a \rangle$, which couple to $\rho$ via neutrino self-interactions.

The situation encountered here is similar to the one that arises in classical kinetics \cite{volpe2013extended, froustey2020neutrino}. There, the Liouville equation is recast in the form of the Bogoliubov--Born--Green--Kirkwood--Yvon (BBGKY) hierarchy. The Boltzmann equation follows from truncating the hierarchy at the first layer. Interactions couple the one- and two-particle distribution functions at a fundamental level, but this lack of closure is remedied by appealing to the hypothesis of molecular chaos. In other words, truncation is effectuated by treating the interactions coupling one- and two-particle distribution functions as collisions: two initially uncorrelated particles briefly exchange energy and momentum, then resume freely propagating, once again uncorrelated.

The neutrino \textbf{quantum kinetic equation (QKE)} generalizes the Boltzmann equation to encompass quantum coherence \cite{sigl1993general, vlasenko2014neutrino}:
\begin{equation}
    i(\partial_t + \bm{\hat{q}} \cdot \partial_{\bm{x}} ) \rho_{\bm{q}} = [H_{\bm{q}}, \rho_{\bm{q}}] + iC_{\bm{q}}, \label{eq:qke}
\end{equation}
with Hamiltonian $H_{\bm{q}}$ and collision operator $iC_{\bm{q}}$. The left-hand side describes the ultrarelativistic advection of neutrinos. Forces, which would appear as terms proportional to $\partial_{\bm{q}}\rho$, are usually assumed to be negligible.

The commutator $[H_{\bm{q}},\rho_{\bm{q}}]$ accounts for \textbf{coherent processes} \cite{wolfenstein1978neutrino, mikheyev1985resonance,  notzold1988neutrino, pantaleone1992neutrino, duan2010collective}:
\begin{equation}
    H_{\bm{q}} = H_{\bm{q}}^\textrm{vac} + H_{\bm{q}}^\textrm{mat} + H_{\bm{q}}^{\nu\nu} = \frac{M^2}{2|\bm{q}|} + \sqrt{2} G_F ( 1 - \bm{\hat{q}} \cdot \bm{v}_m ) L + \sqrt{2} G_F ( D_0 - \bm{\hat{q}} \cdot \bm{D}_1 ). \label{eq:HQKE}
\end{equation}
The matter velocity $\bm{v}_m$ is the local velocity of the astrophysical fluid, not of any particular particle. In the mass basis, $M^2$ is diagonal and has elements $m_i^2$. In the flavor (or weak-interaction) basis, $L$ is diagonal and has elements $n_{\alpha^-} - n_{\alpha_+}$. (We use $\alpha$ to label neutrino flavor and charged-lepton species.) The other matrices are
\begin{equation}
    D_0 = \int \frac{d^3\bm{q}'}{(2\pi)^3} ( \rho_{\bm{q}'} - \bar{\rho}_{\bm{q}'}), ~~~ \bm{D}_1 = \int \frac{d^3\bm{q}'}{(2\pi)^3} \bm{\hat{q}}' ( \rho_{\bm{q}'} - \bar{\rho}_{\bm{q}'}). \label{eq:Dmatrices}
\end{equation}
Here and elsewhere overbars denote antineutrino quantities. In defining the antineutrino density matrix, we adopt the convention $(\bar{\rho})_{ij} = \langle b^\dagger_i b_j \rangle$, where the indices are swapped relative to the neutrino density matrix $(\rho)_{ij} = \langle a^\dagger_j a_i \rangle$ \cite{sigl1993general}. As a result, $\bar{\rho}$ evolves under a formally identical QKE but with Hamiltonian $\bar{H}_{\bm{q}} = - H_{\bm{q}}^\textrm{vac} + H_{\bm{q}}^\textrm{mat} + H_{\bm{q}}^{\nu\nu}$. Some of the literature adopts the convention with same-order indices. In that case, $\bar{\rho} \rightarrow \bar{\rho}^*$ in Eq.~\ref{eq:Dmatrices} and $\bar{H}_{\bm{q}} = (H_{\bm{q}}^\textrm{vac})^* - H_{\bm{q}}^\textrm{mat} - (H_{\bm{q}}^{\nu\nu})^*$. 

The vacuum potential $H^\textrm{vac}$ is intrinsic to neutrinos as elementary particles. The matter potential $H^\textrm{mat}$ and neutrino--neutrino (or self-interaction) potential $H^{\nu\nu}$ arise from coherent interactions of neutrinos with the medium in which they propagate. Various related terms, including \textit{forward scattering}, \textit{refraction}, \textit{effective masses}, and \textit{self-energies}, also refer to the emergence of mean-field potentials from a large number of coherent interactions. $H^{\nu\nu}$ couples $\rho_{\bm{q}}$ to all other $\rho_{\bm{q}'}$ at the same location, with coupling strengths that depend on the factor $1-\bm{\hat{q}}\cdot\bm{\hat{q}}'$. Collective oscillations are made possible by $H^{\nu\nu}$.

\textbf{Incoherent processes} are contained in $iC_{\bm{q}}$. This term consists of matrix-structured Boltzmann collision integrals for all possible momentum- and number-changing interactions involving neutrinos \cite{sigl1993general, blaschke2016neutrino, richers2019neutrino}. Simplifications are frequently used. For example, relaxation-time approximations of the form $iC_{\bm{q}} = \Gamma (\rho^{\textrm{eq}}_{\bm{q}} - \rho_{\bm{q}})$ isolate the tendency of collisions to drive the system toward an equilibrium state $\rho^{\textrm{eq}}_{\bm{q}}$ \cite{johns2021collisional}. As in the classical Boltzmann equation, collisions in the QKE tend to thermalize the number densities $(\rho_{\bm{q}})_{\alpha\alpha}$. Charged-current collisions additionally cause quantum decoherence---decay of $(\rho_{\bm{q}})_{\alpha\beta}$---because they act as measurements of neutrino flavor states. CFI came as a surprise because it defies this expectation: collisions, even charged-current ones, can cause $(\rho_{\bm{q}})_{\alpha\beta}$ to grow exponentially through a feedback process involving the neutrino--neutrino potential.

Derivations of the QKE are manifold. Although the technical frameworks differ, they share certain core features. The scale of neutrino momentum $|\bm{q}|$ is taken to be much greater than the scale of neutrino masses $m$ (the ultrarelativistic assumption), much greater than the scale of the interaction energy $\Sigma \sim G_F n$ (the weak-coupling assumption), and much greater than the scale of gradients $\partial_{\bm{x}}$ in the neutrino field or background medium (the slow-variation assumption) (Fig.~\ref{fig:length_scales}). Furthermore, the influence of multi-particle correlations is folded into collision integrals. For correlations between the neutrinos and their environment, this approximation is based on the idea that any correlation generated during the interaction of a neutrino and an electron, for example, would very quickly be lost by the rethermalization of the electron through electromagnetic interactions. For correlations among the neutrinos, the guiding principle is molecular chaos applied to particles in quantum superpositions.

\subsection{Many-body correlations\label{sec:MB}}

The possibility of significant, molecular-chaos-defying \textbf{many-body correlations} has been explored in a large number of papers using the many-body forward-scattering Hamiltonian
\begin{equation}
    H = \sum_{\bm{q}} \omega_{\bm{q}} \bm{B} \cdot \bm{J}_{\bm{q}} + \frac{\sqrt{2}{G_F}}{V} \sum_{\bm{q},\bm{q}'} ( 1 - \bm{\hat{q}} \cdot \bm{\hat{q}}' ) \bm{J}_{\bm{q}} \cdot \bm{J}_{\bm{q}'}, \label{eq:HMB}
\end{equation}
where $V$ is the quantization volume, $\omega_{\bm{q}} = \delta m^2 / 2|\bm{q}|$ is the vacuum oscillation frequency with mass-squared splitting $\delta m^2$, and $\bm{B} = (0,0,-1)$ is the mass vector written in the mass basis \cite{friedland2003neutrino, bell2003speed, balantekin2006neutrino, patwardhan2022many}. The flavor isospin vector $\bm{J}_{\bm{q}}$ is related to the neutrino creation and annihilation operators via
\begin{equation}
    \bm{J}_{\bm{q}}^+ = a_1^\dagger (\bm{q}) a_2 (\bm{q}), ~~~ \bm{J}_{\bm{q}}^- = a_2^\dagger (\bm{q}) a_1 (\bm{q}), ~~~ \bm{J}_{\bm{q}}^z = [ a_1^\dagger (\bm{q}) a_1 (\bm{q}) - a_2^\dagger (\bm{q}) a_2 (\bm{q}) ] / 2,
\end{equation}
where subscripts denote the two mass states. The Hamiltonian in Eq.~\ref{eq:HMB} reproduces the $iC_{\bm{q}} = 0$ QKE under the mean-field replacement $\bm{J}_{\bm{q}} \rightarrow \langle \bm{J}_{\bm{q}} \rangle$. Calculations reveal substantial entanglement and varying degrees of discrepancy with the kinetic evolution, all the way up to many-body flavor equilibration on a coherent ($\propto G_F^{-1}$) time scale \cite{martin2023equilibration}. On the basis of these findings, collective oscillations have become one of the premier potential applications of quantum computing within the domain of particle and nuclear physics \cite{bauer2023quantum, di2024quantum}.

Several challenges will have to be met as the quantum many-body formulation of neutrino transport continues to develop. One is to find ways of tolerating a very large number of neutrinos given the exponential scaling of the dimension of Hilbert space \cite{xiong2022, roggero2022, martin2022, lacroix2022}. Another originates from the fact that Eq.~\ref{eq:HMB} is actually a requantization of the kinetic Hamiltonian (Eq.~\ref{eq:HQKE}) and, as such, unphysically privileges forward scattering \cite{shalgar2023we, johns2023neutrino}. At the time of writing, only one study has carried out calculations using the complete many-body Hamiltonian \cite{cirigliano2024neutrino}. It reported outcomes qualitatively different from those calculated using Eq.~\ref{eq:HMB}, observing flavor equilibration and momentum thermalization on the same time scale.

A final challenge concerns the issue of what initial states are appropriate for calculations. In particular, the applicability of quantum kinetics may hinge on the existence of localized particles \cite{shalgar2023we, johns2023neutrino}. All fully many-body calculations performed so far have evolved states that are initially product states in the basis of definite-$\bm{q}$ plane waves. It is unclear whether kinetic evolution should be expected for such systems. The quasi-many-body model of Ref.~\cite{kost2024once} highlights this point. Two-body correlations build up during each brief interaction but are then traced out, following the logic of molecular chaos. The evolution in this model exhibits distinct coherent and incoherent contributions, as expected of a kinetic system. An important goal for this research area is to definitively establish the relationship between kinetic evolution and the properties of the underlying quantum state or ensemble.

\section{ESSENTIAL CONCEPTS\label{sec:essential}}

\subsection{Symmetries and conservation laws\label{sec:symmetries}}

Neutrino flavor evolution is often analyzed in terms of \textbf{polarization vectors} $\bm{P}_{\bm{q}} (t, \bm{x})$ and $\bm{\bar{P}}_{\bm{q}} (t, \bm{x})$. In the two-flavor approximation, these are introduced through expansions in Pauli matrices:
\begin{equation}
\rho_{\bm{q}} = (P_{\bm{q},0} + \bm{P}_{\bm{q}}\cdot\bm{\sigma})/2, ~~~~ \bar{\rho}_{\bm{q}} = (\bar{P}_{\bm{q},0} + \bm{\bar{P}}_{\bm{q}}\cdot\bm{\sigma})/2. \label{eq:Pdefns}
\end{equation}
An advantage of polarization vectors is that they draw out the connection between neutrino flavor and angular momentum. Consider the QKE for $\bm{P}_{\bm{q}}$, which follows from plugging the definition above into Eq.~\ref{eq:qke}:
\begin{equation}
    (\partial_t + \bm{\hat{q}} \cdot \partial_{\bm{x}} ) \bm{P}_{\bm{q}} = \bm{H}_{\bm{q}} \times \bm{P}_{\bm{q}} + \bm{C}_{\bm{q}}. \label{eq:Pqke}
\end{equation}
The Hamiltonian causes $\bm{P}_{\bm{q}}$ to precess around $\bm{H}_{\bm{q}}$. This motion is analogous to the Larmor precession of a magnetic moment, except that the precession is in flavor space. From Eq.~\ref{eq:HQKE},
\begin{equation}
    \bm{H}_{\bm{q}} = \omega_{\bm{q}} \bm{B} + \sqrt{2} G_F(1-\bm{\hat{q}}\cdot\bm{v}_m) \bm{L} + \sqrt{2} G_F ( \bm{D}_0 - \bm{\hat{q}} \cdot \bm{\vec{D}}_1 ), \label{eq:Hvec}
\end{equation}
with $\omega_{\bm{q}} = \delta m^2 / 2 |\bm{q}|$ and $\delta m^2=m_2^2-m_1^2$. The signs of $\delta m^2$ and $\omega_{\bm{q}}$ reflect the neutrino mass ordering. All flavor-space vectors are defined through expansions similar to Eq.~\ref{eq:Pdefns}. We write $\bm{\vec{D}}_1$ with an arrow to distinguish it from $\bm{D}_1$ in Eq.~\ref{eq:Dmatrices}. By our chosen $\bar{\rho}$ convention, $\bm{\bar{H}}_{\bm{q}}$ is identical to $\bm{H}_{\bm{q}}$ but with $\omega\bm{B} \rightarrow -\omega\bm{B}$. We will stick with the flavor basis from this point on, so that $\bm{z}$ always represents the flavor axis. Assuming the muon and tau populations are negligible, the charged-lepton vector is $\bm{L} = n_e \bm{z}$, with net electron rest number density $n_e$. The mass vector is $\bm{B} = (\sin 2\theta)\bm{x}-(\cos 2\theta)\bm{z}$, where $\theta$ is the vacuum mixing angle.

Polarization vectors can be defined for three-flavor mixing using Gell-Mann matrices, though the visual intuition is more tenuous because they are eight-dimensional \cite{dasgupta2008collective}. We will continue to assume two flavors for simplicity. Much of the literature makes the same assumption, anticipating that the extension to three flavors will be relatively straightforward once the oscillation problem can be solved for two.

As with any physical system, conservation laws are of paramount importance. Take $\bm{C}_{\bm{q}} = 0$ for the moment. Then $P_{\bm{q},0}$, the number of neutrinos with momentum $\bm{q}$ regardless of flavor, and $|\bm{P}_{\bm{q}}|$, the flavor polarization magnitude, are conserved along neutrino trajectories. This statement implies conservation of \textbf{entropy} under collisionless evolution because
\begin{equation}
    S_{\bm{q}} = - \textrm{Tr} \left[ \rho_{\bm{q}} \log \rho_{\bm{q}} + (1-\rho_{\bm{q}}) \log (1-\rho_{\bm{q}}) \right] \label{eq:entropy}
\end{equation}
is a function only of $P_{\bm{q},0}$ and $|\bm{P}_{\bm{q}}|$. Boltzmann collision integrals respect the second law of thermodynamics, $dS/dt \geq 0$, with $S$ being summed over all particles \cite{sigl1993general}. When a neutrino interacts with its environment, or with another neutrino, some amount of correlation is generated. Quantum kinetics discards this information in its description of interactions (Sec.~\ref{sec:processes}), rendering collisions entropy-increasing and irreversible.

Oscillations and forward scattering generate entropy only in a coarse-grained sense. The term \textbf{kinematic decoherence} labels the underlying mechanism, in which polarization vectors that are nearby in phase space become misaligned in flavor space \cite{raffelt2007self}. When coarse-graining is performed by averaging over small phase-space regions, the misalignment results in depolarization. As a simple example, consider averaging two polarization vectors $\bm{P}_{\bm{q}}$ and $\bm{P}_{\bm{q}'}$. The result satisfies $|\langle \bm{P} \rangle| \leq \langle | \bm{P} | \rangle$, with equality if and only if $\bm{P}_{\bm{q}}$ and $\bm{P}_{\bm{q}'}$ are aligned. From the concavity of Eq.~\ref{eq:entropy} it follows that $S[\langle \rho \rangle] \geq \langle S[\rho] \rangle$. Misalignment leads to coarse-grained depolarization, which leads to coarse-grained entropy increase \cite{johns2023thermodynamics}. The process is kinematic in the sense that it stems from the collisionless propagation of neutrinos. To what extent the second law of thermodynamics applies to coarse-grained coherent evolution is currently unknown.

Another important quantity is the system energy, which has contributions from the neutrino masses, momenta, and coherent interactions. The overwhelming part comes from the momenta because neutrinos are ultrarelativistic and weakly interacting. But for the same reason, this part of the energy is virtually uninfluenced by the flavor-space dynamics. It can be useful to focus on the energy associated with the latter. The \textbf{polarization energy} is
\begin{align}
    U(t) = \frac{1}{2} \int \frac{d^3\bm{x} d^3\bm{q}}{(2\pi)^3} \left( \left( \bm{H}_{\bm{q}}^{\textrm{vac}} + \bm{H}_{\bm{q}}^{\textrm{mat}} + \frac{1}{2} \bm{H}_{\bm{q}}^{\nu\nu} \right) \cdot \bm{P}_{\bm{q}} - \left( \bm{\bar{H}}_{\bm{q}}^{\textrm{vac}} + \bm{\bar{H}}_{\bm{q}}^{\textrm{mat}} + \frac{1}{2} \bm{\bar{H}}_{\bm{q}}^{\nu\nu} \right) \cdot \bm{\bar{P}}_{\bm{q}} \right),
\end{align}
with integrals running over the entire system \cite{duan2006collective, raffelt2007self}. $U$ is conserved in a collisionless, spatially homogeneous system. Any individual $U_{\bm{q}}$ is able to evolve because neutrino--neutrino forward scattering facilitates the exchange of polarization energy between neutrinos at different momenta. In an inhomogeneous setting, $U$ conservation is broken because neutrino advection nontrivially couples the internal (flavor-space) and external (phase-space) dynamics \cite{fiorillo2024inhomogeneous}.

One last notable object characterizing a neutrino system is the \textbf{total difference vector}
\begin{equation}
        \boldsymbol{\Delta}(t) = \int \frac{d^3\bm{x}d^3\bm{q}}{(2\pi)^3}\left( \bm{P}_{\bm{q}} - \bm{\bar{P}}_{\bm{q}} \right). \label{eq:Delta}
\end{equation}
Its projection along $\bm{z}$ is related to the difference of electron-flavor lepton number and muon-flavor lepton number (the eLN$-\mu$LN, as it is sometimes stylized) through $2\Delta_z = ( N_{\nu_e} - N_{\bar{\nu}_e}) - ( N_{\nu_\mu} -N_{\bar{\nu}_\mu})$, where $N_{\nu_\alpha}$ is the total number of neutrinos with flavor $\alpha$. In the limit of zero vacuum mixing, the flavored lepton numbers $L_{\nu_e}$ and $L_{\nu_\mu}$ are individually conserved, and therefore total lepton number $L = L_{\nu_e} + L_{\nu_\mu}$ and lepton-number difference $2\Delta_z = L_{\nu_e} - L_{\nu_\mu}$ are independent invariants. The constraints on the polarizations that flow from conservation of $L_{\nu_e}$ and $L_{\nu_\mu}$ are contained in the condition $\Delta_z = \textrm{constant}$. With nonzero vacuum mixing, $L_{\nu_e}$ and $L_{\nu_\mu}$ are not strictly conserved, nor is $\Delta_z$. 

A defining feature of the oscillation problem is that calculations are only tractable for relatively simple models. Simplification, however, is an art. Often it introduces symmetries and additional invariants that can artificially stabilize neutrino flavor states and otherwise fundamentally change the dynamics, as strikingly demonstrated by the slow \cite{hannestad2006self, duan2007analysis, xiong2023symmetry, fiorillo2023slow}, fast \cite{johns2020neutrino, padilla2022neutrino, xiong2023symmetry, fiorillo2023slow}, and collisional \cite{johns2023collisional} flavor pendula. To borrow an adage from Nigel Goldenfeld: \textit{Symmetry is the enemy of instability}. The potential for insight is balanced by the risk of unwarranted extrapolation. With all models, it is wise to ask which features are general, which are informatively specific, and which are just irrelevant.

\subsection{Adiabaticity\label{sec:adiabaticity}}

Following the classic \textbf{bulb model}, suppose we approximate the neutrinos in a CCSN as streaming collisionlessly outward from a sharp emitting surface (the \textit{neutrinosphere}). Assume that everything---the neutrinosphere, the neutrino radiation field, the background of other particles---is spherically symmetric. Assume also that the astrophysical environment changes so slowly that we can take it to be static. It then makes sense to search for steady states of Eq.~\ref{eq:Pqke} by integrating
\begin{equation}
    \cos\vartheta \frac{d\bm{P}_{\omega,\vartheta}}{dr} = \bm{H}_{\omega,\vartheta}\times\bm{P}_{\omega,\vartheta} \label{eq:bulbmulti}
\end{equation}
in radius $r$ outward from the neutrinosphere. By symmetry, the momentum label $\bm{q}$ of a neutrino reduces to its vacuum oscillation frequency $\omega$ (a proxy for $|\bm{q}|$) and the emission angle $\vartheta$. The majority of the literature on collective oscillations prior to around 2016 took Eq.~\ref{eq:bulbmulti} as a starting point. In many cases, further simplifications were adopted. For example, in the \textit{single-angle approximation}, all neutrinos are assumed to have $\vartheta$-independent radial evolution. The equation of motion in this case becomes simply
\begin{equation}
    \frac{d\bm{P}_{\omega}}{dr} = \bm{H}_{\omega} \times \bm{P}_{\omega}. \label{eq:bulbsingle}
\end{equation}
Details such as the explicit forms of $\bm{H}_{\omega,\vartheta}$ and $\bm{H}_\omega$ can be found in Refs.~\cite{duan2010collective, mirizzi2016supernova}. The relevant point here is that we arrive at essentially the same equation one would solve for solar neutrinos, apart from the inclusion of the neutrino--neutrino potential. In approaching the oscillation problem in CCSNe and NSMs, a reasonable strategy is to start from the oscillation problem in the Sun, which is already solved, and systematically add in elements of the problem until reaching the full Eq.~\ref{eq:Pqke}. This has more or less been the path followed historically.

Flavor transformation in the Sun is to a large extent explained by the principle of \textbf{adiabaticity}. Take the viewpoint of a single neutrino traveling out from the Sun's core. It sees a slowly varying, nearly time-constant external potential due to the background particles. It stands to reason that the flavor state $| \psi (t) \rangle$ should obey the adiabatic theorem, stating that the overlap $|\langle \nu_I(t) | \psi (t)\rangle|$ with the $I$th instantaneous energy eigenstate is constant. In the limit of high neutrino energy, $\omega$ is small and the Hamiltonian is dominated by the potential arising from ambient electrons. As a result, a high-energy $\nu_e$ created in the solar core is nearly in an energy eigenstate upon production. By the adiabatic theorem, it remains in that eigenstate all the way out into vacuum, even as the flavor composition of the state changes with declining $n_e$. This mechanism underlies the MSW effect.

In CCSNe and NSMs, the issue of adiabaticity is deeper and more nuanced. MSW resonances are generally expected to be adiabatic, though disruption is possible due to sharp interfaces (\textit{e.g.}, the supernova shock passing through the resonance region) \cite{schirato2002connection} or fluctuations (\textit{e.g.}, hydrodynamic turbulence in the resonance region) \cite{duan2009neutrino}. MSW conversion is also influenced by the neutrino--neutrino potential if it happens that this potential is large where the vacuum and matter potentials nearly cancel \cite{duan2008flavor,dasgupta2008spectral}.

Moving beyond MSW resonances, the phenomenon of spectral swaps can be explained by appealing to the notion of \textit{nonlinear} adiabaticity \cite{duan2006simulation, duan2006coherent, duan2007analysis, fogli2007collective, raffelt2007self2, raffelt2007adiabaticity, dasgupta2009multiple}. Suppose that $\bm{\hat{P}}_\omega = s_\omega \bm{\hat{H}}_\omega$ at all radii, with constant $s_\omega = \pm 1$ in some rotating frame \cite{raffelt2007self2, raffelt2007adiabaticity}. This requires that self-consistency conditions be satisfied because each $\bm{\hat{H}}_\omega$ depends on $\bm{P}_{\omega'}$ for all $\omega' \neq \omega$. Similar conditions arise in the context of Bardeen--Cooper--Schrieffer (BCS) pairing, where the ground state must likewise be determined self-consistently due to the presence of mean-field interactions \cite{pehlivan2017spectral}. For neutrinos, the self-consistent solutions are known as \textbf{pure-precession states} because all $\bm{P}_\omega$ undergo precession in the lab frame with a single frequency $\Omega$ around $\bm{B}$. This behavior can be contrasted with vacuum oscillations, where each $\bm{P}_\omega$ precesses with its own frequency $\omega$ around $\bm{B}$. In the limit $n_\nu \rightarrow \infty$, pure-precession states exhibit synchronized oscillations, with all $\bm{P}_\omega$ aligned with one another. As $n_\nu$ decreases, neutrinos evolve adiabatically through a sequence of pure-precession states, ending with all neutrinos in mass states. Spectral swaps form in the process because $\bm{P}_\omega$ asymptotically aligns with $\bm{B}$ for some energies and anti-aligns for others. 

One of the puzzles to come out of the bulb model was that adiabaticity sometimes accounted for the numerically observed behavior and sometimes conflicted with it. The effort to determine the conditions under which oscillations are adiabatic motivated the formalization of linear stability analysis as it applies to neutrino flavor (Sec.~\ref{sec:stability}). Ultimately, the systematic study of flavor instabilities did not merely clarify the evolution that had been observed numerically. It undermined the use of the bulb model as a way to predict flavor transformation in the first place. The reason was already alluded to in Sec.~\ref{sec:symmetries}: the stringent assumptions of the bulb model restrict the evolution in unacceptable ways, as evidenced by the presence of symmetry-breaking flavor instabilities. The story for NSMs parallels the one for CCSNe. In simple merger models, adiabaticity explains the numerically observed evolution through matter--neutrino resonances \cite{malkus2012neutrino, malkus2014matter, wu2015physics, malkus2015symmetric, zhu2016matter, frensel2016neutrino}. These models appear to be compromised by oversimplification as well \cite{shalgar2017multi, padilla-gay:2024symmetry}. 

It is now believed that the full nature of the QKE must generally be taken into account. Collisional and coherent regimes of evolution cannot typically be sequestered from one another, nor can the identity of the QKE as a partial differential equation typically be reduced to that of an ordinary differential equation. There is no consensus yet as to where this leaves the adiabatic phenomena of spectral swaps and matter--neutrino resonant transformation, which were premised on simplifications now viewed with suspicion.

\subsection{Instability\label{sec:stability}}

Flavor instabilities signal deviations between the classical and quantum kinetic theories of neutrinos. In \textbf{linear stability analysis}, the QKE is linearized by assuming that $|\rho^{e\mu} | \ll | \rho^{ee} - \rho^{\mu\mu}|$, or that all $\bm{P}_{\bm{q}}$ are nearly aligned or anti-aligned with $\bm{z}$ \cite{banerjee2011linearized, izaguirre2017fast}. Linear stability analysis imagines that flavor mixing slightly perturbs neutrinos away from weak-interaction eigenstates, then asks whether the perturbations remain small or grow large. Collective modes of the form
\begin{equation}
    \rho_{\bm{q}}^{e\mu} (t, \bm{x}) = Q_{\bm{q}}(\bm{K}) \exp (-i\Omega (\bm{K}) t + i \bm{K} \cdot \bm{x})
\end{equation}
are considered. The factor $Q_{\bm{q}}(\bm{K})$ is an eigenfunction of the linearized system. Although flavor-coherence nonlinearity is no longer present, collectivity still is: $\rho_{\bm{q}}^{e\mu}$ decouples from $\rho_{\bm{q}'}^{e\mu}$ but not from $\rho_{\bm{q}'}^{ee}$. A collective mode is one in which essentially all momenta participate due to the coupling of $\bm{q}$ to $\bm{q}'$. There is no coupling between modes at different wave vectors $\bm{K}$ and $\bm{K}'$ because the background is nearly homogeneous by assumption. Once $\bm{K}$ is specified, the eigensystem is solved to produce $Q_{\bm{q}}(\bm{K})$ and eigenvalue $\Omega (\bm{K})$.

An equation relating $\Omega$ to $\bm{K}$ is a \textbf{dispersion relation} for flavor-coherence waves propagating on a homogeneous background. The analytic properties of dispersion relations and their dynamical significance are a rich subject interwoven with connections to the physics of fluids and plasmas \cite{capozzi2017fast, capozzi2019fast, yi2019dispersion, fiorillo2024theory, fiorillo2024theoryfast, fiorillo2024theoryslow, fiorillo2025theoryslow, fiorillo2025collective}.

There are in fact multiple solutions $\Omega$ for any given $\bm{K}$. In many instances, only the maximum $\textrm{Im}(\Omega)$ is really of interest, because it reveals whether perturbations grow or not. If there exists a collective mode with $\textrm{Im}(\Omega) > 0$ for some $\bm{K}$, then the system is unstable. Linear stability analysis has most importantly been used for the purpose of identifying flavor instabilities in astrophysical simulations (Sec.~\ref{sec:prevalence}). The procedure involves post-processing: a simulation is conducted \textit{without} oscillations, and then the output is probed with the question, ``Should this simulation have been carried out with flavor mixing included?'' Linear stability analysis produces a dispersion relation $\Omega(\bm{K})$ for each time step and spatial grid point. A single grid point in the simulation is understood as representing an infinite homogeneous medium in which the flavor-coherence waves are defined. To find out when and where flavor instabilities compromise the reliability of neutrino classical kinetics, the maximum value of $\textrm{Im}(\Omega)$ is calculated throughout the spatial domain of the simulation for select snapshots in time.

By itself, this procedure does not shed much light on what gives rise to flavor instabilities. Deeper understanding comes from considering the linearized evolution of a neutrino system in certain idealized limits. Here the common three-type classification of instabilities enters (Sec.~\ref{sec:prevalence}). The first class is FFI \cite{sawyer2009multiangle, sawyer2016neutrino}. By definition, an instability is fast if it occurs when all vacuum oscillation frequencies $\omega$ and all collision rates $\Gamma$ are set to zero in the dispersion relation. Growth rates can be as high as $\mathcal{O}(\mu)$, where $\mu = \sqrt{2} G_F n_\nu$. The category of SFI includes all the collisionless instabilities that disappear at $\omega = 0$ \cite{kostelecky1993neutrino, duan2006collective}. The speediest among them have $\mathcal{O}(\sqrt{\mu\omega})$ growth rates. CFI rounds out the taxonomy \cite{johns2021collisional}. As the name implies, CFI disappears at $\Gamma = 0$. Its growth rates reach up to $\mathcal{O}(\sqrt{\mu\Gamma})$. It is important to stress that these rates are highly approximate and are not even the correct scaling in all cases. For example, FFI identified in a simulation usually has a rate quite a bit smaller than $\mu$, and CFI more typically has a rate roughly proportional to $\Gamma$.

Occasionally one encounters other instability types, such as \textit{bipolar}, \textit{multi-azimuthal-angle}, and \textit{multi-zenith-angle}. These terms are somewhat more historical, though they can still be useful depending on the context. For the newcomer to collective oscillations, the important thing to know is that they do not refer to distinct phenomena not covered here. 

Any classification of instabilities is bound to be somewhat arbitrary. However, the division into fast, slow, and collisional has a compelling connection to the (non)conservation of $\boldsymbol{\Delta}$ (Eq.~\ref{eq:Delta}). Ignoring the matter potential, Eqs.~\ref{eq:Pqke} and \ref{eq:Hvec} imply
\begin{equation}
    \frac{d\boldsymbol{\Delta}}{dt} = \int \frac{d^3\bm{x}d^3\bm{q}}{(2\pi)^3} \left( \omega_{\bm{q}} \bm{B} \times \left( \bm{P}_{\bm{q}} + \bm{\bar{P}}_{\bm{q}} \right) + \bm{C}^{-}_{\bm{q}} \right), \label{eq:dDdt}
\end{equation}
where $\bm{C}^{-}_{\bm{q}} = \bm{C}_{\bm{q}} - \bm{\bar{C}}_{\bm{q}}$. SFI and CFI, which respectively have the requisite conditions $\omega_{\bm{q}} \neq 0$ and $\bm{C}_{\bm{q}} \neq 0$, are associated with two distinct ways of breaking $\boldsymbol{\Delta}$ conservation. The particular significance of the collisional asymmetry is suggested by the appearance of $\bm{C}_{\bm{q}}^-$ in Eq.~\ref{eq:dDdt}. In fact, introducing a separate symbol $\bar{\omega}_{\bm{q}} = - \omega_{\bm{q}}$ for antineutrinos shows that the first mechanism of $\boldsymbol{\Delta}$ nonconservation arises specifically from the vacuum asymmetry $\omega^{-}_{\bm{q}} = \omega_{\bm{q}} - \bar{\omega}_{\bm{q}}$, since $\omega_{\bm{q}}^{+} = \omega_{\bm q} + \bar\omega_{\bm q} = 0$.

In a system with $\omega_{\bm{q}} = 0$ and $\bm{C}_{\bm{q}} = 0$, $\boldsymbol{\Delta}$ is an invariant. Any (necessarily fast) instabilities in such a system must operate within this constraint. Define
\begin{equation}
    \bm{D}_{\bm{\hat{q}}} (t,\bm{x}) = \int \frac{d|\bm{q}||\bm{q}|^2}{2\pi^2}\left( \bm{P}_{\bm{q}} - \bm{\bar{P}}_{\bm{q}} \right)
\end{equation}
and assume that the system initially has the same value of $D_{\bm{\hat{q}},z}$ for all $\bm{x}$, with small $\bm{x}$-dependent perturbations in the transverse vectors $\bm{D}_{\bm{\hat{q}},T}$. For FFI to occur, the system must initially have an \textbf{angular crossing}: $D_{\bm{\hat{q}},z}$ must pass through zero at some $\bm{\hat{q}}$. The \textit{necessity} of an angular crossing can be demonstrated using linear stability analysis \cite{morinaga2022fast, dasgupta2022collective} or inferred from the conservation of $\Delta_z$ using the relation $\Delta_z = V \int \frac{d\bm{\hat{q}}}{4\pi} D_{\bm{\hat{q}},z}$ with system volume $V$ \cite{johns2024ergodicity, fiorillo2024theory}. The proof that angular crossings are \textit{sufficient} for FFI is significantly more technical \cite{morinaga2022fast, fiorillo2024theory}.

The concept of \textbf{self-induced resonance} helps illuminate the physics of flavor instabilities \cite{raffelt2008self, johns2020neutrino}.  We will explain the basic idea using the equation of motion $\partial_t \bm{P}_{\bm{q}} = \bm{H}_{\bm{q}} \times \bm{P}_{\bm{q}}$. Each polarization vector $\bm{P}_{\bm{q}}$ attempts to precess around its Hamiltonian $\bm{H}_{\bm{q}}$, but $\bm{H}_{\bm{q}}$ itself is evolving. If the relative azimuthal angle between these vectors evolves rapidly, then the term $\bm{H}_{\bm{q},T} \times \bm{P}_{\bm{q},T}$, which drives the evolution of $P_{\bm{q},z}$, is expected to average to zero over time. However, if $\bm{H}_{\bm{q},T}$ and $\bm{P}_{\bm{q},T}$ maintain a fixed relative phase over a long enough duration---in other words, if there is a resonance between the frequencies---then a secular change in $P_{\bm{q},z}$ can accumulate. This idea extends to spatially inhomogeneous systems, where individual neutrinos can additionally be resonant with $\bm{K} \neq 0$ collective waves \cite{fiorillo2024theory}. This is essentially the same particle--wave resonance mechanism that is central to Landau damping in plasmas and other physical systems.

\begin{figure}
    \centering
    \begin{minipage}{0.24\linewidth}
    \includegraphics[width=\linewidth]{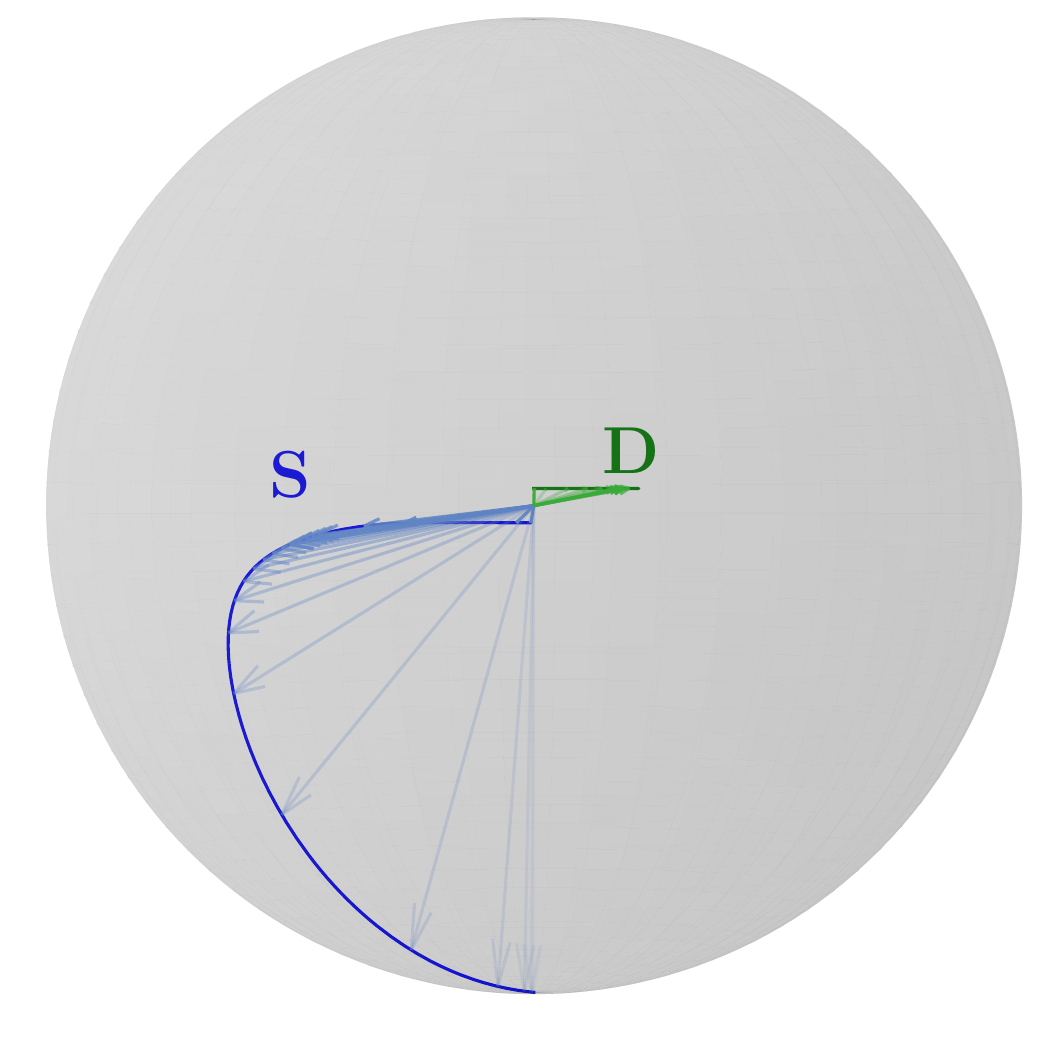}
    \end{minipage}
    \begin{minipage}{0.37\linewidth}
    \includegraphics[width=.97\linewidth, trim=2cm 2cm 3cm 2cm, clip]{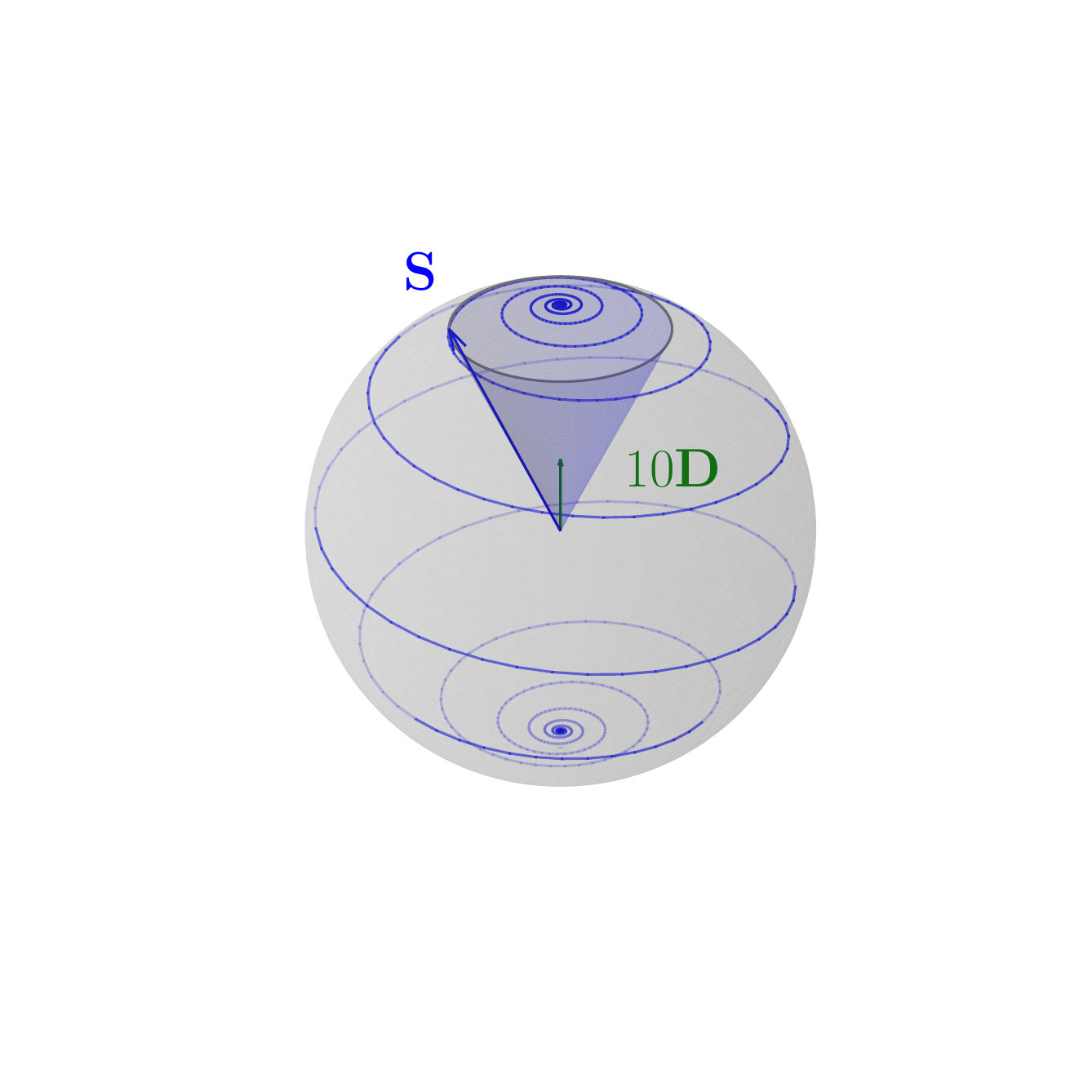}
    \end{minipage}
    \begin{minipage}{0.37\linewidth}
    \includegraphics[width=\linewidth, trim=4cm 2cm 0cm 1cm, clip]{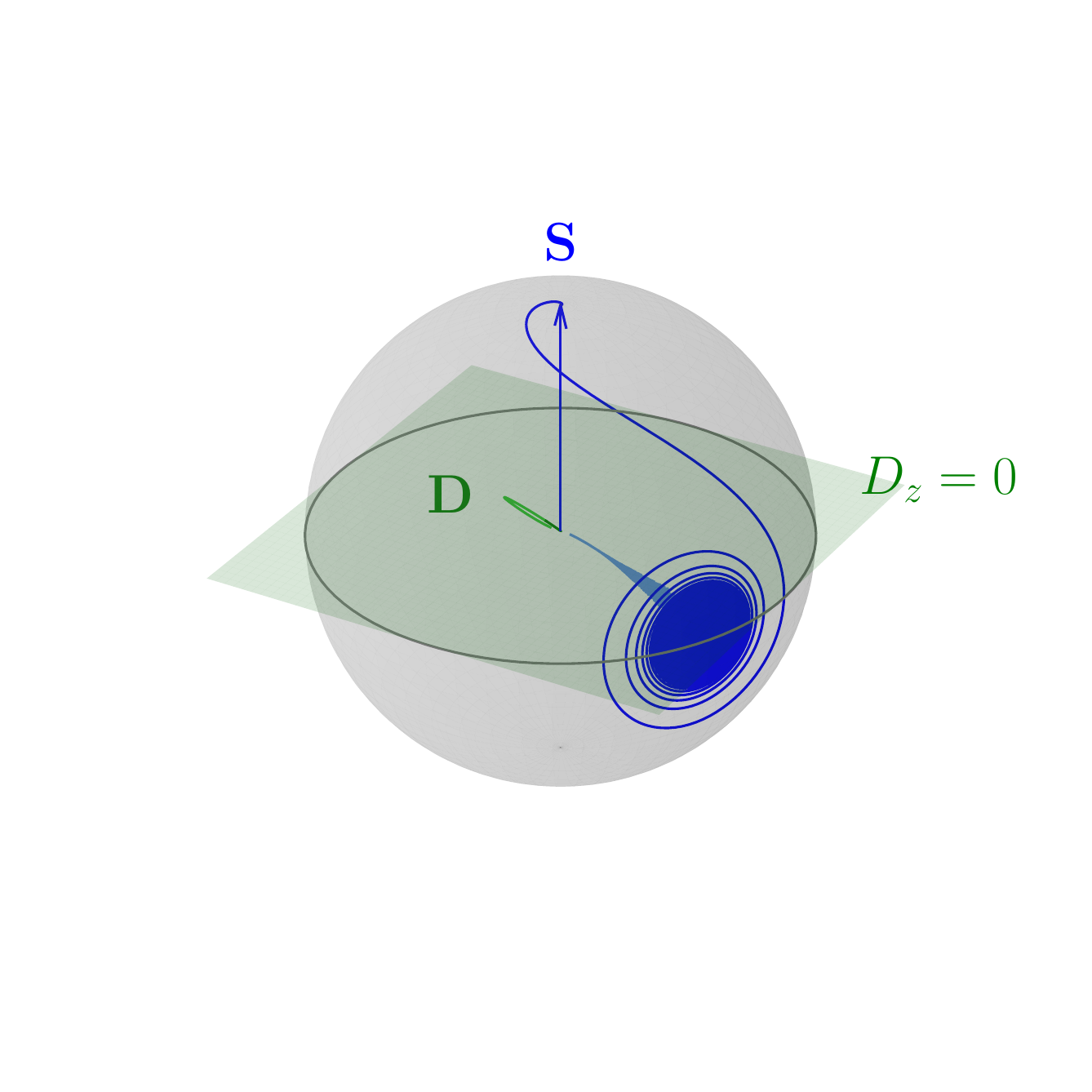}
    \end{minipage}
    \caption{Essential mechanisms of CFI based on the analysis of a simple model (Eq.~\ref{eq:CFI} with $\Gamma^- > 0$). \textit{Left:} When $\bm{S}$ and $\bm{D}$ are initially in opposite directions, they remain anti-aligned. $\bm{S}$ tilts toward the plane transverse to the flavor axis $\bm{z}$ as the transverse part $\bm{D}_T$ grows. \textit{Center:} When $\bm{S}$ and $\bm{D}$ are initially in the same direction, $\bm{S}$ precesses around $\bm{D}$ while small fluctuations in $\bm{D}$ cause $\bm{S}$ to tilt away from $\bm{z}$. \textit{Right:} When $\bm{D} \approx 0$ initially, $\bm{S}$ plunges down toward the resonance as $\bm{D}_T$ grows. The data plotted are from Ref. \cite{johns2023collisional}. See text for details.}
    \label{fig:cfi}
\end{figure}

Resonance functions somewhat differently in CFI \cite{johns2021collisional, johns2023collisional}. Consider the idealized system governed by $\partial_t \bm{P} = \sqrt{2} G_F \bm{D}\times\bm{P} - \Gamma\bm{P}_T$ and $\partial_t \bm{\bar{P}} = \sqrt{2} G_F \bm{D}\times\bm{\bar{P}} - \bar{\Gamma}\bm{\bar{P}}_T$, which are coupled through $\bm{D} = \bm{P} - \bm{\bar{P}}$. Introducing $\Gamma^\pm = (\Gamma \pm \bar{\Gamma})/2$, we rewrite these equations in terms of the difference vector $\bm{D}$ and sum vector $\bm{S} = \bm{P} + \bm{\bar{P}}$:
\begin{equation}
    \frac{d \bm{S}}{dt} = \sqrt{2}G_F\bm{D}\times\bm{S} - \Gamma^+ \bm{S}_T - \Gamma^-\bm{D}_T, ~~~~ \frac{d \bm{D}}{dt} = - \Gamma^+ \bm{D}_T - \Gamma^- \bm{S}_T. \label{eq:CFI}
\end{equation}
While $\Gamma^+$ simply causes flavor decoherence, $\Gamma^-$ promotes (anti-)alignment of $\bm{S}_T$ and $\bm{D}_T$. If phase-locking is maintained, $\bm{D}_T$ grows exponentially and drags $\bm{S}$ toward the transverse plane (left image in Fig.~\ref{fig:cfi}). This behavior is characteristic of one regime of CFI. In another regime, the $\bm{D}\times\bm{S}$ term rapidly dephases $\bm{S}$ and $\bm{D}$ and causes CFI to be far from resonance (middle image). In the intermediate regime, $\bm{D}$ is approximately in the transverse plane from the beginning and $\bm{S}$ swiftly (rate $\propto \sqrt{\mu\Gamma}$) falls into resonance with it (right image). The last scenario is called \textit{resonance-like CFI} \cite{xiong2023collisional, liu2023systematic}. It is indeed resonant in the sense that the potential nearly vanishes along the flavor axis ($\sqrt{2}G_F D_z \approx 0$), but in the sense of phase evolution it marks the boundary between resonant and nonresonant.

\subsection{Matter effects\label{sec:matter}}

One might expect the matter deep inside a CCSN or NSM to move the flavor mixing very far from resonance and thus have a strongly suppressive effect on flavor conversion \cite{wolfenstein1979neutrino}. The issue turns out not to be so simple. Although it is customary to talk about \textit{the} matter effect for neutrinos propagating in the Sun or Earth, in more extreme environments it is more appropriate to speak of \textbf{matter effects}, plural.

The fact that matter does not always have the naively expected influence can be seen by adopting a rotating frame that eliminates the matter potential \cite{duan2006collective, hannestad2006self}. Consider again the QKE $\partial_t \bm{P}_{\bm{q}} = \bm{H}_{\bm{q}} \times \bm{P}_{\bm{q}}$, which contains a term $\lambda \bm{z} \times \bm{P}_{\bm{q}}$ arising from neutrino--electron forward scattering, with $\lambda = \sqrt{2} G_F n_e$. In a frame rotating about $\bm{z}$ with frequency $\lambda$, the highlighted term drops out. The penalty paid for this simplification is that the mass vector $\bm{B}$ becomes time-dependent. But if $\lambda$ is very large, then $\bm{B}$, upon averaging over time, should appear to point very nearly along $\bm{z}$. This suggests that the matter potential can be ignored if the vacuum mixing angle $\theta$ is replaced by a suitably suppressed value. A related observation in linear stability analysis is that $\lambda$ can be absorbed into a shift of $\textrm{Re}(\Omega)$ with no change to $\textrm{Im}(\Omega)$. From either perspective, the conclusion is that $\lambda$ does not necessarily nullify collective mechanisms of flavor conversion.

On the other hand, sometimes it does. The phenomenon of \textbf{multi-angle matter suppression} dramatically alters flavor conversion in the bulb model and other setups where instabilities develop spatially \cite{esteban2008role, chakraborty2011no}. Consider the QKE in the multi-angle bulb model (Eq.~\ref{eq:bulbmulti}). After dividing through by $\cos\vartheta$, the matter term becomes $(\lambda / \cos\vartheta)\bm{L} \times \bm{P}_{\omega,\vartheta}$, which cannot be eliminated for all trajectories simultaneously with a single rotating-frame transformation. Intuitively, neutrinos propagating outward on more tangential paths have to traverse more matter in order to cover the same radial distance as neutrinos that are moving out more radially. 

Clearly there is some validity to the rotating-frame argument, and it is certainly a boon not to have $\lambda$ determining the time step in numerical calculations. But in light of the many advances of recent years, a comprehensive reassessment of matter effects is now warranted \cite{sigl2022, xiong2022evolution, fiorillo2024theoryslow, shalgar2025neutrino}. A thorough analysis will need to account for the matter-\textit{flux} potential as well. Although this term has almost always been neglected, it is not obviously insignificant \cite{padilla2021fast}. The structure of the term puts it on par with the anisotropic part of the neutrino--neutrino potential: compare the factors of $\bm{\hat{q}}\cdot\bm{v}_m$ and $\bm{\hat{q}}\cdot\bm{D}_1$ in Eq.~\ref{eq:HQKE}. The matter-flux potential can be competitive with the neutrino--neutrino potential even though $|\bm{v}_m| \ll 1$.

\subsection{Turbulence and equilibrium\label{sec:equilibrium}}

In an earlier era of research on collective oscillations, it was hypothesized that neutrino--neutrino forward scattering might produce neutrino distributions of such fine structure---``\textit{developing modes of higher and higher multipolarity}''---that flavor equipartition would effectively be achieved \cite{sawyer2005speed}. Years later, this suggestion seems rather prescient. It contains two ideas that are now considered essential. An abundance of numerical evidence shows that flavor instabilities, at least SFI and FFI, generally coincide with a cascade in phase space, whereby neutrino flavor distributions form features on progressively smaller scales \cite{raffelt2007self, mangano2014damping, johns2020fast, bhattacharyya2020late, bhattacharyya2020fast, wu2021collective}. This phenomenon is a form of turbulence facilitated by the nonlinearity of neutrino refraction in dense environments. A fluctuating or quasi-steady asymptotic state is reached, which is sometimes referred to as \textit{flavor equilibrium} (though it is not a state of \textit{equipartition}, with an equal number of neutrinos in each flavor, except under special circumstances).

\begin{figure}
    \centering
    \begin{minipage}{0.44\linewidth}
    \includegraphics[width=\linewidth]{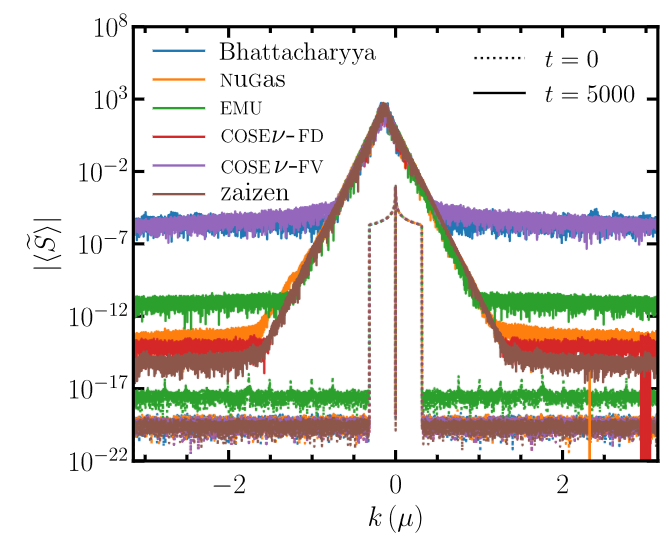}
    \end{minipage}
    \begin{minipage}{0.49\linewidth}
    \includegraphics[width=\linewidth]{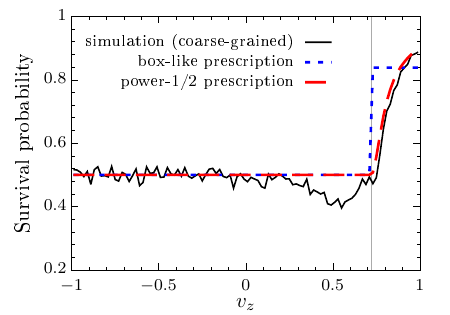}
    \end{minipage}
    \caption{\textit{Left:} Angle-averaged Fourier power spectrum of the off-diagonal (flavor-coherent) element $\rho^{e\mu}$ of the neutrino density matrix, showing late-time exponential scaling in flavor-wave turbulence. Results from several quantum-kinetics codes are compared. Adapted with permission from Richers et al. (2022) \cite{richers2022code}. \textit{Right:} Spatially averaged post-FFI $\nu_e$ survival probability, comparing two empirical formulas to the exact numerical result. Adapted with permission from Xiong et al. (2023) \cite{xiong2023evaluating}.
    }
    \label{fig:pee}
\end{figure}

The physics of \textbf{flavor-wave turbulence} is poorly understood. Linear stability analysis describes the evolution of flavor waves in the regime of small flavor coherence. Nonlinear solutions are known for individual flavor waves on static, homogeneous backgrounds \cite{duan2021flavor}. However, once a spectrum of flavor waves becomes significantly populated, nonlinearity couples the waves to one another. This mechanism evidently establishes a fluctuating, chaotic balance across spatial scales, often with an exponential scaling as a function of $|\bm{K}|$ \cite{richers2021neutrino, richers2022code, urquilla2024chaos} (left panel of Fig.~\ref{fig:pee}). Unlike in hydrodynamic turbulence, the balance is not between large-scale driving and small-scale dissipation, but is instead negotiated by the interactions among flavor waves at all $\bm{K}$. A quantitative theory does not exist yet, and even the foregoing qualitative remarks should be viewed as tentative.

An ongoing strand of research is devoted to the characterization of post-instability \textbf{asymptotic states} (Sec.~\ref{sec:coarse}). The effort to date has focused almost exclusively on FFI. The most important finding is that angular crossings, the precondition for FFI, are absent from the spatially averaged flavor distributions in the asymptotic state. No theoretically derived mapping from initial to asymptotic state is known. However, empirical formulas motivated by numerical calculations have been proposed \cite{zaizen2022simple, xiong2023evaluating, richers2024asymptoic, george2024evolution}. They all make the assumption that the asymptotic eLN$-\mu$LN (see below Eq.~\ref{eq:Delta}) vanishes on one angular domain. They take different forms on the other domain. All of them enforce the conservation of $\Delta_z$, the eLN$-\mu$LN integrated over angle. Two examples are highlighted and compared to the exact numerical result in the right panel of Fig.~\ref{fig:pee}. Other studies have explored the possibility of using machine learning to predict the asymptotic coarse-grained angular moments from the initial ones supplied by a moment-based CCSN or NSM simulation \cite{abbar2023physics, abbar2024application, richers2024asymptoic}.

The mappings above were all developed for calculations with periodic boundary conditions. Conservation laws and asymptotic states are different when Dirichlet boundary conditions are adopted \cite{zaizen2023characterizing}. Models with neutrinos emitted from the boundaries show asymptotic phenomena (\textit{e.g.}, \textit{fast flavor swaps}) with coherent structure in $\bm{x}$ \cite{zaizen2024fast}. The issue of what boundary conditions are most appropriate for subgrid models has not yet been resolved (Sec.~\ref{sec:coarse}).

Not much is known about asymptotic states resulting from the other types of instabilities. CFI exhibits interesting spectral features due to the dependence of collision rates on neutrino energy \cite{lin2023collision}. A notable example is the \textit{collisional flavor swap} \cite{kato2023collisional, zaizen2025spectral}. The lack of separation between the CFI scale and the macroscopic astrophysical scale makes it challenging to estimate the effects of instability even if CFI asymptotic states are available. Analyses suggest that significant conversion of $\nu_e$ and $\bar{\nu}_e$ into the heavy-lepton flavors could occur, but it is also possible that significant growth of CFI is prevented by the advection of neutrinos out of the decoupling region \cite{johns2021collisional, johns2022collisional, xiong2022evolution, shalgar2023neutrinos, shalgar2024neutrino}. 

The precise relevance of post-instability asymptotic states for CCSNe and NSMs is contested. If the time scale for flavor to equilibrate is much shorter than the time scale for astrophysical conditions to change, then the types of initial conditions that have been used to calculate asymptotic states---namely, flavor states that are far from stability---are not expected to appear in astrophysical settings \cite{johns2024subgrid}. Taken to the extreme, the separation of the equilibration and astrophysical scales motivates the hypothesis that oscillations maintain a state of local equilibrium, with flavor configurations evolving quasistatically from one equilibrium to the next \cite{johns2023thermodynamics}. Whether local-equilibrium evolution does occur in CCSNe and NSMs, and whether it is accurately approximated by a sequence of post-instability asymptotic states, are open questions (Sec.~\ref{sec:coarse}).

\section{STRATEGIES FOR SOLVING THE OSCILLATION PROBLEM\label{sec:strategies}}

\subsection{Attenuated Hamiltonians}

The oscillation problem calls for transport methods that accurately approximate neutrino flavor dynamics. Resolving oscillations in full detail is a nonstarter because of the mesoscopic scales involved. State-of-the-art CCSN simulations typically employ resolutions of $\sim\mathcal{O}(100)$~m and $\mathcal{O}(100)~$ns at best. If the oscillation scales of $\mathcal{O}(1)$~cm and $\mathcal{O}(1)$~ns are any indication, a solution of the QKE without approximation would necessitate orders of magnitude more memory and computation time.

Solving the QKE in anything like a realistic model of a CCSN or NSM will require shrewd techniques to minimize computing costs as much as possible. One proposal is to apply an artificial \textbf{attenuation factor} to the strength of the neutrino--neutrino potential, rescaling that part (and possibly other parts) of the Hamiltonian by a factor $\xi$ \cite{nagakura2022general, nagakura2022time, xiong2022evolution}. Attenuation is predicated on the notion that the influence of flavor mixing becomes independent of the strength of the potential in the limit that the potential is large. Convergent solutions of the QKE on static CCSN/NSM background profiles have been obtained with $\xi$ typically in the range of $10^{-5}$ to $10^{-3}$.

The argument has also been made that the QKE can be accurately solved using macroscopic spatial resolution even \textit{without} attenuation because collisions and advection naturally smooth out gradients of the neutrino flavor field \cite{shalgar2022neutrino, shalgar2023neutrino}. This assertion is controversial, however \cite{nagakura2025resolution}.

Global solutions of the QKE on static backgrounds, whether using attenuation factors or not, have produced a number of findings: alteration of neutrino decoupling \cite{shalgar2022neutrino}, the influence of advection on flavor instabilities \cite{shalgar2019neutrino, xiong2022evolution}, varying degrees of flavor equipartition \cite{padillagay2021multi,nagakura2021constructing,nagakura2023bhatnagar, xiong2024fast,shalgar2023neutrino, shalgar2024neutrino}, reduction of neutrino heating in the supernova gain region along with enhancement of total neutrino luminosity \cite{nagakura2023roles, xiong2024fast}, the elimination of angular crossings in quasisteady states \cite{nagakura2022time, nagakura2022connecting, xiong2024fast}, and fast flavor swaps in a post-merger environment \cite{nagakura2023global}. They have also been used to demonstrate agreement with subgrid treatments of oscillations based on FFI asymptotic states \cite{xiong2024robust}, a topic that we return to in Sec.~\ref{sec:coarse}.

\subsection{Quantum closures}

Another technique involves resolution not in $t$ or $\bm{x}$ but in $\bm{\hat{q}}$. Because tracking the evolution of full phase-space distribution functions is costly, many radiation-hydrodynamic simulations implement schemes in which only the leading angular moments are retained. When expressed in terms of moments, the neutrino transport equation turns into an infinite tower of coupled equations. Moment-based simulations truncate the tower using various closures, which prescribe the unevolved moments in terms of the moments whose evolution is explicitly solved for. 

Quantum kinetics, like neutrino transport without oscillations, can be recast in terms of moments \cite{strack2005generalized}. A problem that arises is that some oscillation calculations, particularly in the context of the bulb model (Sec.~\ref{sec:adiabaticity}), require $\mathcal{O}(1000)$ angle bins to achieve convergence \cite{duan2008simulating}. Fortunately for moment-based quantum kinetics, this stringent resolution requirement does not appear to be an intrinsic feature of neutrino flavor conversion. It is instead related to \textit{spurious instabilities} attributable to the use of discrete angle bins \cite{sarikas2012spurious}. Spurious instabilities disappear in the moment formalism \cite{morinaga2018linear, johns2020neutrino}.

The first moment solutions of the QKE were nevertheless not especially encouraging \cite{duan2014multipole}. Another type of spurious evolution is possible, unrelated to spurious instabilities: truncation error at the smallest angular scales can propagate up to large scales and have an outsize impact on the overall flavor evolution \cite{johns2020fast}. An important caveat, however, is that these studies used trivial closures, setting all unevolved moments to zero. 

More recent work has sought to formulate suitable \textbf{quantum closures} (\textit{i.e.}, angular-moment closures that include information about flavor coherence) for moment-based quantum kinetics. Initial studies using ad hoc closure prescriptions found enough success in replicating the results of linear stability analysis and nonlinear evolution to warrant further investigation \cite{myers2022neutrino, grohs2023neutrino, grohs2023twomoment, froustey2023neutrino}. The challenges now are to construct generally applicable and theoretically sound quantum closures \cite{kneller2024quantum, froustey2024quantum} and to ensure full compatibility between quantum moments and the computational methods of astrophysical simulations \cite{grohs2025advection}.

\subsection{Coarse-grained theories\label{sec:coarse}}

Other approaches do not solve the QKE per se but instead adopt a coarse-grained treatment of neutrino oscillations. This category includes effective classical transport \cite{li2021neutrino, xiong2024robust}, the Bhatnagar--Gross--Krook (BGK) subgrid model \cite{nagakura2023bhatnagar}, miscidynamics \cite{johns2023thermodynamics, johns2024subgrid}, quasilinear theory \cite{fiorillo2024fast, fiorillo2025collective}, and quasi-homogeneous analysis \cite{liu2024quasi}.

\begin{figure}
    \centering
    \includegraphics[width=0.75\linewidth]{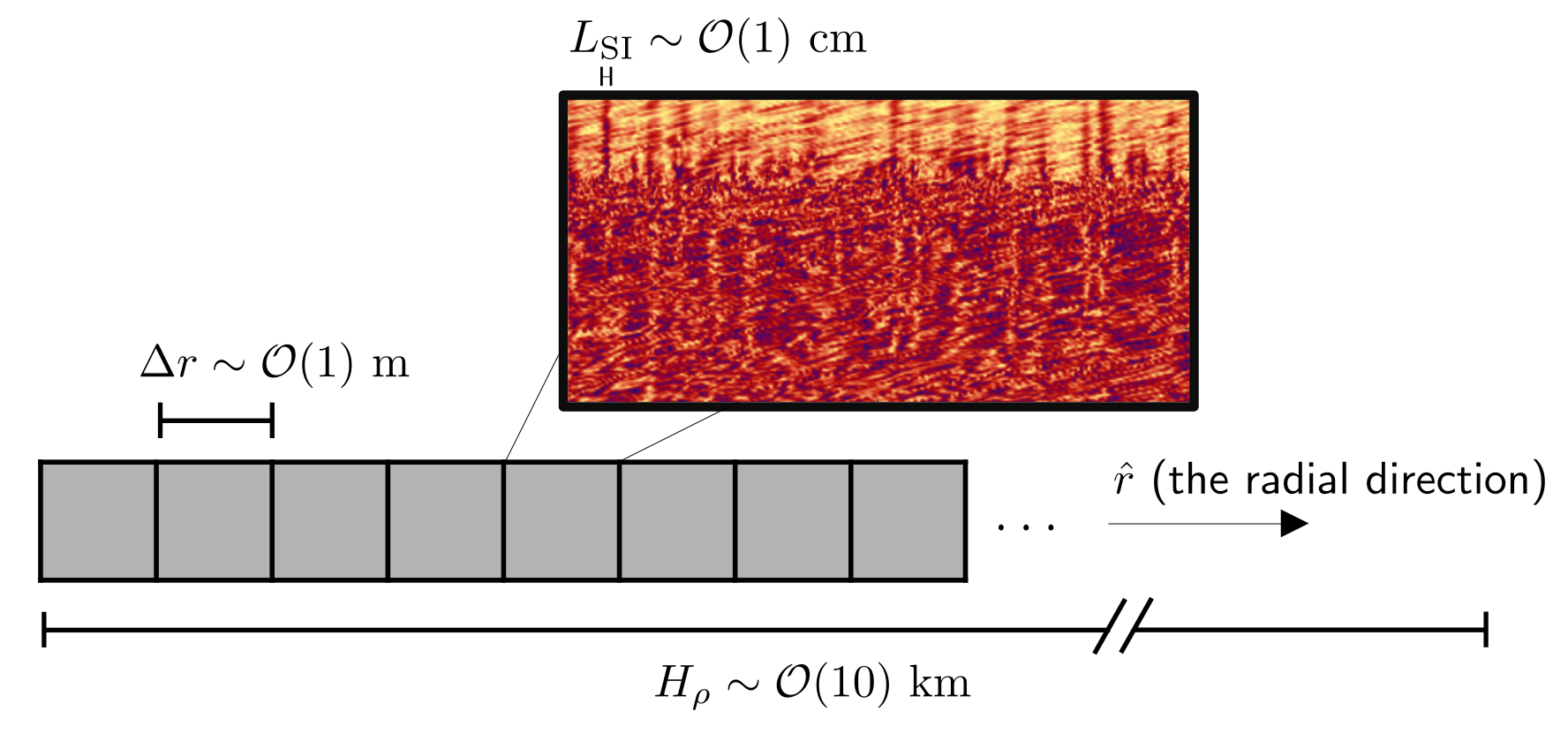}
    \caption{Schematic illustration of subgrid approaches to the oscillation problem. $L_{\rm SI}$ and $H_\rho$ are the self-interaction scale and density scale height of Fig.~\ref{fig:length_scales}, respectively. $\Delta r$ represents a possible coarse-graining scale, which draws the distinction between grid-level and subgrid physics.}
    \label{fig:periodic_box_cartoon}
\end{figure}

The first two are inspired by the numerical observation that FFI rapidly brings flavor distributions to asymptotic states (Sec.~\ref{sec:equilibrium}). We imagine dividing up a CCSN or NSM into boxes with side length $\Delta r$ in between the mesoscopic oscillation scale $\sim L_{\rm SI}$ and the macroscopic astrophysical scale $\sim H_\rho$  (Fig.~\ref{fig:periodic_box_cartoon}). The evolution in each of these small regions is then approximated by the evolution that occurs if the box is isolated from its surroundings and flavor instabilities are allowed to develop. An asymptotic-state subgrid model must address two key issues: how to impose the asymptotic states, and how to calculate them.

In \textbf{effective classical transport}, asymptotic states are applied using a two-step procedure \cite{li2021neutrino, xiong2024robust}. The first step advances the simulation forward in time while ignoring oscillations. We know from linear stability analysis of simulation data that this step frequently produces unstable flavor distributions (Sec.~\ref{sec:prevalence}). Those distributions are then fed in as the initial conditions of a subgrid QKE calculation in order to produce asymptotic states. (Or, alternatively, the mapping from initial to asymptotic states is approximated, as we discuss below.) In the second step, flavor is instantaneously redistributed within each unstable region to match the corresponding asymptotic state. The two-step sequence then repeats. 

Effective classical transport is used in most of the recent estimates of oscillation effects (Sec.~\ref{sec:effects}). Clearly the procedure is computationally practical. Moreover, it appears to quantitatively agree with global quantum kinetic solutions of fast flavor conversion on static backgrounds \cite{xiong2024robust}. Quantum kinetic solutions show that spatially coarse-grained density matrices have nearly vanishing flavor coherence. This is an important mark of consistency with effective classical transport, for which asymptotic states are necessarily flavor-diagonal.

An alternative asymptotic-state method replaces $[H_{\bm{q}},\rho_{\bm{q}}]$ in the QKE with the \textbf{quantum BGK approximation} $\Gamma^{\textrm{a}} (\rho^{\textrm{a}}_{\bm{q}} - \rho_{\bm{q}})$, where $\Gamma^{\textrm{a}}$ is the relaxation rate and $\rho^{\textrm{a}}_{\bm{q}}$ is the asymptotic state \cite{nagakura2023bhatnagar}. This quantum relaxation-time approximation can be compared with the collisional relaxation-time approximation in Sec.~\ref{sec:processes}. As in the two-step procedure, oscillations are assumed to bring neutrinos toward local asymptotic states. The intent is to allow for finite rates of relaxation toward these states, rather than imposing instantaneous relaxation. Qualitative agreement has been found when comparing BGK and quantum kinetic solutions in a model with neutrino injection from the inner boundary of a radial shell \cite{nagakura2023bhatnagar}.

Regarding the calculation of asymptotic states, the optimal choice would be to run the actual isolated-box QKE calculation for every unstable region. Because this is computationally prohibitive, approximate options are being explored (Sec.~\ref{sec:equilibrium}). Among these are various empirical formulas motivated by QKE calculations \cite{zaizen2022simple, xiong2023evaluating, richers2024asymptoic, george2024evolution}, including the two shown in the right panel of Fig.~\ref{fig:pee}, which were adopted in Ref.~\cite{xiong2024robust}. Other methods use machine learning to develop the mapping from unstable flavor configurations to asymptotic states \cite{abbar2023physics, abbar2024application, richers2024asymptoic}. Irrespective of which option is selected, boundary conditions must be chosen whenever a small region is plucked out and isolated from the rest of the astrophysical environment. Scale separation of the kind illustrated in Fig.~\ref{fig:periodic_box_cartoon} motivates periodic boundary conditions, as adopted in the references above. Asymptotic states differ when Dirichlet boundary conditions are used instead \cite{cornelius2023perturbing, zaizen2023characterizing, zaizen2024fast}.

If flavor instabilities instantaneously lead to asymptotic states, as assumed in the two-step procedure, then strongly unstable states should not materialize in the first place \cite{johns2024subgrid}. This observation calls into question whether asymptotic-state methods apply scale separation in a self-consistent manner. Quantum kinetic calculations of fast flavor conversion do indeed show that oscillations continuously act to eliminate angular crossings \cite{fiorillo2024fast, xiong2024robust}, so that at no point during the evolution does the system resemble the initial states that have thus far been used to calculate asymptotic states. Furthermore, effective classical transport is not capable of accounting for all possible behavior because there can be situations where subgrid quantum coherence is essential for the evolution \cite{johns2024subgrid, liu2024quasi}. This issue may become more pronounced when considering how to generalize asymptotic-state methods beyond FFI to encompass other phenomena like SFI and CFI \cite{zaizen2025spectral}. Yet the fact remains that in many instances these methods are quantitatively successful.

Coarse-grained theories of neutrino transport are in a state of flux. For the approaches relying on asymptotic states, effective classical transport and the BGK subgrid model, issues need to be addressed regarding their self-consistency and generalizability. In addition to asymptotic-state methods, various approaches are being developed based on \textbf{local mixing equilibrium} \cite{johns2023thermodynamics, johns2024subgrid}, \textbf{quasilinear theory} \cite{fiorillo2024fast, fiorillo2025collective}, and \textbf{quasi-homogeneous analysis} \cite{liu2024quasi}. We anticipate continued rapid progress on this front.

\section{PATHS FORWARD}

The research area covered in this review orbits around a single, simple question: How do neutrinos oscillate in extreme stellar environments? The answer to the same question for \textit{ordinary} stellar environments like the Sun is settled textbook material. As we have seen, the oscillation problem in CCSNe and NSMs demands a much deeper understanding of neutrino transport theory---and although much progress has been made, more innovations are needed.

In closing, we highlight a few specific questions that may help guide the way toward a solution to the oscillation problem:

\begin{itemize}

\item A clear picture of the prevalence of FFI in simulations of CCSNe and NSMs is taking shape. Estimates of the effects of oscillations on these astrophysical events are largely motivated by FFI. What are the prevalence and effects of \textbf{CFI and SFI}?

\item Many state-of-the-art radiation-hydrodynamic simulations employ moment methods. Can \textbf{quantum closures} be used to seamlessly and faithfully integrate oscillations into these frameworks?

\item The most straightforward approach to the oscillation problem is to replace, in a simulation, the classical neutrino Boltzmann equation with the QKE. The mesoscopic scales associated with oscillations make this approach computationally infeasible. Is there a reliable \textbf{coarse-grained approximation} of the QKE that we might implement instead?

\item Calculations based on quantum many-body theory produce results that are inconsistent with quantum kinetics. Do these discrepancies disappear in more realistic \textbf{quantum many-body models} of neutrino transport?

\end{itemize}

Of course, these are only a small selection of the many questions worth pursuing on the rich topic of neutrino oscillations in CCSNe and NSMs.

\section*{ACKNOWLEDGMENTS} 
We acknowledge Sajad Abbar, Ryuichiro Akaho, Jakob Ehring, Julien Froustey, and Zewei Xiong for providing figures and simulation data. We are grateful to Basudeb Dasgupta, Damiano Fiorillo, Gabriel Mart\'{i}nez-Pinedo, Georg Raffelt, G\"{u}nter Sigl, Irene Tamborra, and Zewei Xiong for comments on the manuscript.
LJ is supported by a Feynman Fellowship through LANL LDRD project number 20230788PRD1. SR is supported by the National Science Foundation under Grant No. 2412683. 
MRW  acknowledges support of the National Science and Technology Council, Taiwan under Grant No.~111-2628-M-001-003-MY4, the Academia Sinica under Project No. AS-IV-114-M04, and the Physics Division of the National Center for Theoretical Sciences, Taiwan. 

%




\bibliographystyle{ar-style5} 
\bibliography{references}

\begin{thebibliography}{194}
\expandafter\ifx\csname natexlab\endcsname\relax\def\natexlab#1{#1}\fi

\bibitem{janka2025long}
Janka H, et~al.
\newblock \textit{arXiv preprint arXiv:2502.14836}  (2025)

\bibitem{duan2010collective}
Duan H, Fuller GM, Qian YZ.
\newblock \textit{Annu. Rev. Nucl. Part. Sci.} 60(1):569--594 (2010)

\bibitem{mirizzi2016supernova}
Mirizzi A, Tamborra I, Janka HT, Saviano N, Scholberg K, et~al.
\newblock \textit{Riv. Nuovo Cimento} 39:1--112 (2016)

\bibitem{volpe2024neutrinos}
Volpe MC.
\newblock \textit{Rev. Mod. Phys.} 96(2):025004 (2024)

\bibitem{fischer2024neutrinos}
Fischer T, Guo G, Langanke K, Martinez-Pinedo G, Qian YZ, Wu MR.
\newblock \textit{Prog. Part. Nucl. Phys.} 137:104107 (2024)

\bibitem{chakraborty2016collective}
Chakraborty S, Hansen R, Izaguirre I, Raffelt G.
\newblock \textit{Nucl. Phys. B} 908:366--381 (2016)

\bibitem{tamborra2021new}
Tamborra I, Shalgar S.
\newblock \textit{Annu. Rev. Nucl. Part. Sci.} 71(1):165--188 (2021)

\bibitem{richers2023fast}
Richers S, Sen M.
\newblock In \textit{Handbook of Nuclear Physics}. Springer,  3771--3787 (2023)

\bibitem{patwardhan2022many}
Patwardhan AV, Cervia MJ, Rrapaj E, Siwach P, Balantekin AB.
\newblock In \textit{Handbook of Nuclear Physics}. Springer,  1--16 (2022)

\bibitem{capozzi2022neutrino}
Capozzi F, Saviano N.
\newblock \textit{Universe} 8(2):94 (2022)

\bibitem{tamborra2024neutrinos}
Tamborra I.
\newblock \textit{arXiv preprint arXiv:2412.09699}  (2024)

\bibitem{wolfenstein1978neutrino}
Wolfenstein L.
\newblock \textit{Phys. Rev. D} 17(9):2369--2374 (1978)

\bibitem{mikheyev1985resonance}
Mikheyev SP, Smirnov AY.
\newblock \textit{Sov. J. Nucl. Phys.} 42(6):913--917 (1985)

\bibitem{notzold1988neutrino}
N{\"o}tzold D, Raffelt G.
\newblock \textit{Nucl. Phys. B} 307(4):924--936 (1988)

\bibitem{pantaleone1992neutrino}
Pantaleone J.
\newblock \textit{Phys. Lett. B} 287(1-3):128--132 (1992)

\bibitem{kersten2016decoherence}
Kersten J, Smirnov AY.
\newblock \textit{Eur. Phys. J. C} 76:1--20 (2016)

\bibitem{akhmedov2017collective}
Akhmedov E, Kopp J, Lindner M.
\newblock \textit{J. Cosmol. Astropart. Phys.} 2017(09):017 (2017)

\bibitem{banerjee2011linearized}
Banerjee A, Dighe A, Raffelt G.
\newblock \textit{Phys. Rev. D} 84(5):053013 (2011)

\bibitem{izaguirre2017fast}
Izaguirre I, Raffelt G, Tamborra I.
\newblock \textit{Phys. Rev. Lett.} 118(2):021101 (2017)

\bibitem{sawyer2009multiangle}
Sawyer RF.
\newblock \textit{Phys. Rev. D} 79(10):105003 (2009)

\bibitem{sawyer2016neutrino}
Sawyer RF.
\newblock \textit{Phys. Rev. Lett.} 116(8):081101 (2016)

\bibitem{wu2017fast}
Wu MR, Tamborra I.
\newblock \textit{Phys. Rev. D} 95(10):103007 (2017)

\bibitem{abbar2019occurrence}
Abbar S, Duan H, Sumiyoshi K, Takiwaki T, Volpe MC.
\newblock \textit{Phys. Rev. D} 100(4):043004 (2019)

\bibitem{delfan2019linear}
Delfan~Azari M, Yamada S, Morinaga T, Iwakami W, Okawa H, et~al.
\newblock \textit{Phys. Rev. D} 99(10):103011 (2019)

\bibitem{morinaga2020fast}
Morinaga T, Nagakura H, Kato C, Yamada S.
\newblock \textit{Phys. Rev. Res.} 2(1):012046 (2020)

\bibitem{abbar2021characteristics}
Abbar S, Capozzi F, Glas R, Janka HT, Tamborra I.
\newblock \textit{Phys. Rev. D} 103(6):063033 (2021)

\bibitem{nagakura2021where}
Nagakura H, Burrows A, Johns L, Fuller GM.
\newblock \textit{Phys. Rev. D} 104(8):083025 (2021)

\bibitem{li2021neutrino}
Li X, Siegel DM.
\newblock \textit{Phys. Rev. Lett.} 126(25):251101 (2021)

\bibitem{fernandez2022fast}
Fern\'andez R, Richers S, Mulyk N, Fahlman S.
\newblock \textit{Phys. Rev. D} 106(10):103003 (2022)

\bibitem{johns2021collisional}
Johns L.
\newblock \textit{Phys. Rev. Lett.} 130(19):191001 (2023)

\bibitem{xiong2022evolution}
Xiong Z, Wu MR, Mart\'\i{}nez-Pinedo G, Fischer T, George M, et~al.
\newblock \textit{Phys. Rev. D} 107(8):083016 (2023)

\bibitem{xiong2023collisional}
Xiong Z, Johns L, Wu MR, Duan H.
\newblock \textit{Phys. Rev. D} 108(8):083002 (2023)

\bibitem{liu2023universality}
Liu J, Akaho R, Ito A, Nagakura H, Zaizen M, Yamada S.
\newblock \textit{Phys. Rev. D} 108(12):123024 (2023)

\bibitem{akaho2023collisional}
Akaho R, Liu J, Nagakura H, Zaizen M, Yamada S.
\newblock \textit{Phys. Rev. D} 109(2):023012 (2024)

\bibitem{shalgar2023neutrinos}
Shalgar S, Tamborra I.
\newblock \textit{Phys. Rev. D} 109(10):103011 (2024)

\bibitem{froustey2024neutrino}
Froustey J, Richers S, Grohs E, Flynn SD, Foucart F, et~al.
\newblock \textit{Phys. Rev. D} 109(4):043046 (2024)

\bibitem{liu2023systematic}
Liu J, Zaizen M, Yamada S.
\newblock \textit{Phys. Rev. D} 107(12):123011 (2023)

\bibitem{kostelecky1993neutrino}
Kosteleck{\`y} VA, Samuel S.
\newblock \textit{Phys. Lett. B} 318(1):127--133 (1993)

\bibitem{duan2006collective}
Duan H, Fuller GM, Qian YZ.
\newblock \textit{Phys. Rev. D} 74(12):123004 (2006)

\bibitem{neto2023energy}
Neto PD, Tamborra I, Shalgar S.
\newblock \textit{arXiv preprint arXiv:2312.06556}  (2023)

\bibitem{shalgar2024neutrino}
Shalgar S, Tamborra I.
\newblock \textit{J. Cosmol. Astropart. Phys.} 09:021 (2024)

\bibitem{fiorillo2024theoryslow}
Fiorillo DFG, Raffelt GG.
\newblock \textit{arXiv preprint arXiv:2412.02747}  (2024)

\bibitem{fiorillo2025theoryslow}
Fiorillo DFG, Raffelt GG.
\newblock \textit{arXiv preprint arXiv:2501.16423}  (2025)

\bibitem{johns2021fast}
Johns L, Nagakura H.
\newblock \textit{Phys. Rev. D} 103(12):123012 (2021)

\bibitem{nagakura2021constructing}
Nagakura H, Johns L.
\newblock \textit{Phys. Rev. D} 103(12):123025 (2021)

\bibitem{mukhopadhyay2024time}
Mukhopadhyay P, Miller J, McLaughlin GC.
\newblock \textit{arXiv preprint arXiv:2404.17938}  (2024)

\bibitem{capozzi2020mu}
Capozzi F, Chakraborty M, Chakraborty S, Sen M.
\newblock \textit{Phys. Rev. Lett.} 125(25):251801 (2020)

\bibitem{capozzi2021fast}
Capozzi F, Abbar S, Bollig R, Janka HT.
\newblock \textit{Phys. Rev. D} 103(6):063013 (2021)

\bibitem{shalgar2021three}
Shalgar S, Tamborra I.
\newblock \textit{Phys. Rev. D} 104(2):023011 (2021)

\bibitem{liu2024muon}
Liu J, Nagakura H, Akaho R, Ito A, Zaizen M, et~al.
\newblock \textit{Phys. Rev. D} 110(4):043039 (2024)

\bibitem{ehring2023fastneutrino}
Ehring J, Abbar S, Janka HT, Raffelt G, Tamborra I.
\newblock \textit{Phys. Rev. Lett.} 131(6):061401 (2023)

\bibitem{just2022fast}
Just O, Abbar S, Wu MR, Tamborra I, Janka HT, Capozzi F.
\newblock \textit{Phys. Rev. D} 105(8):083024 (2022)

\bibitem{abbar2024using}
Abbar S, Volpe MC.
\newblock \textit{arXiv preprint arXiv:2401.10851}  (2024)

\bibitem{schirato2002connection}
Schirato RC, Fuller GM.
\newblock \textit{arXiv preprint astro-ph/0205390}  (2002)

\bibitem{duan2009neutrino}
Duan H, Kneller JP.
\newblock \textit{J. Phys. G} 36(11):113201 (2009)

\bibitem{duan2008flavor}
Duan H, Fuller GM, Carlson J, Qian YZ.
\newblock \textit{Phys. Rev. Lett.} 100(2):021101 (2008)

\bibitem{duan2006simulation}
Duan H, Fuller GM, Carlson J, Qian YZ.
\newblock \textit{Phys. Rev. D} 74(10):105014 (2006)

\bibitem{duan2006coherent}
Duan H, Fuller GM, Carlson J, Qian YZ.
\newblock \textit{Phys. Rev. Lett.} 97(24):241101 (2006)

\bibitem{malkus2012neutrino}
Malkus A, Kneller JP, McLaughlin GC, Surman R.
\newblock \textit{Phys. Rev. D} 86:085015 (2012)

\bibitem{malkus2014matter}
Malkus A, Friedland A, McLaughlin GC.
\newblock \textit{arXiv preprint arXiv:1403.5797}  (2014)

\bibitem{duan2011influence}
Duan H, Friedland A, McLaughlin GC, Surman R.
\newblock \textit{J. Phys. G} 38(3):035201 (2011)

\bibitem{choubey2010signatures}
Choubey S, Dasgupta B, Dighe A, Mirizzi A.
\newblock \textit{arXiv preprint arXiv:1008.0308}  (2010)

\bibitem{dasgupta2012role}
Dasgupta B, O’Connor EP, Ott CD.
\newblock \textit{Phys. Rev. D} 85(6):065008 (2012)

\bibitem{wu2015effects}
Wu MR, Qian YZ, Mart\'{\i}nez-Pinedo G, Fischer T, Huther L.
\newblock \textit{Phys. Rev. D} 91(6):065016 (2015)

\bibitem{sasaki2017possible}
Sasaki H, Kajino T, Takiwaki T, Hayakawa T, Balantekin AB, Pehlivan Y.
\newblock \textit{Phys. Rev. D} 96(4):043013 (2017)

\bibitem{ko2020neutrino}
Ko H, Cheoun MK, Ha E, Kusakabe M, Hayakawa T, et~al.
\newblock \textit{Astrophys. J. Lett.} 891(1):L24 (2020)

\bibitem{stapleford2020coupling}
Stapleford CJ, Fr\"ohlich C, Kneller JP.
\newblock \textit{Phys. Rev. D} 102(8):081301 (2020)

\bibitem{ehring2023fast}
Ehring J, Abbar S, Janka HT, Raffelt G, Tamborra I.
\newblock \textit{Phys. Rev. D} 107(10):103034 (2023)

\bibitem{nagakura2023roles}
Nagakura H.
\newblock \textit{Phys. Rev. Lett.} 130(21):211401 (2023)

\bibitem{mori2025three}
Mori K, Takiwaki T, Kotake K, Horiuchi S.
\newblock \textit{arXiv preprint arXiv:2501.15256}  (2025)

\bibitem{wang2025effect}
Wang T, Burrows A.
\newblock \textit{arXiv preprint arXiv:2503.04896}  (2025)

\bibitem{wu2017imprints}
Wu MR, Tamborra I, Just O, Janka HT.
\newblock \textit{Phys. Rev. D} 96(12):123015 (2017)

\bibitem{xiong2020potential}
Xiong Z, Sieverding A, Sen M, Qian YZ.
\newblock \textit{Astrophys. J.} 900(2):144 (2020)

\bibitem{fujimoto2023explosive}
Fujimoto Si, Nagakura H.
\newblock \textit{Mon. Not. R. Astron. Soc.} 519(2):2623--2629 (2023)

\bibitem{qiu2025neutrino}
Qiu Y, Radice D, Richers S, Bhattacharyya M.
\newblock \textit{arXiv preprint arXiv:2503.11758}  (2025)

\bibitem{lund2025angle}
Lund KA, Mukhopadhyay P, Miller JM, McLaughlin G.
\newblock \textit{arXiv preprint arXiv:2503.23727}  (2025)

\bibitem{george2020fast}
George M, Wu MR, Tamborra I, Ardevol-Pulpillo R, Janka HT.
\newblock \textit{Phys. Rev. D} 102(10):103015 (2020)

\bibitem{gava2009dynamical}
Gava J, Kneller J, Volpe C, McLaughlin GC.
\newblock \textit{Phys. Rev. Lett.} 103(7):071101 (2009)

\bibitem{ehring2024gravitational}
Ehring J, Abbar S, Janka H, Raffelt G, Nakamura K, et~al.
\newblock \textit{arXiv preprint arXiv:2412.02750}  (2024)

\bibitem{volpe2013extended}
Volpe C, V{\"a}{\"a}n{\"a}nen D, Espinoza C.
\newblock \textit{Phys. Rev. D} 87(11):113010 (2013)

\bibitem{serreau2014neutrino}
Serreau J, Volpe C.
\newblock \textit{Phys. Rev. D} 90(12):125040 (2014)

\bibitem{cirigliano2015new}
Cirigliano V, Fuller GM, Vlasenko A.
\newblock \textit{Phys. Lett. B} 747:27--35 (2015)

\bibitem{kartavtsev2015neutrino}
Kartavtsev A, Raffelt G, Vogel H.
\newblock \textit{Phys. Rev. D} 91(12):125020 (2015)

\bibitem{fiorillo2024collective}
Fiorillo DFG, Raffelt GG, Sigl G.
\newblock \textit{Phys. Rev. D} 109(4):043031 (2024)

\bibitem{raffelt1993non}
Raffelt G, Sigl G, Stodolsky L.
\newblock \textit{Phys. Rev. Lett.} 70(16):2363 (1993)

\bibitem{sigl1993general}
Sigl G, Raffelt G.
\newblock \textit{Nucl. Phys. B} 406(1-2):423--451 (1993)

\bibitem{stirner2018liouville}
Stirner T, Sigl G, Raffelt G.
\newblock \textit{J. Cosmol. Astropart. Phys.} 2018(05):016 (2018)

\bibitem{froustey2020neutrino}
Froustey J, Pitrou C, Volpe MC.
\newblock \textit{J. Cosmol. Astropart. Phys.} 2020(12):015 (2020)

\bibitem{vlasenko2014neutrino}
Vlasenko A, Fuller GM, Cirigliano V.
\newblock \textit{Phys. Rev. D} 89(10):105004 (2014)

\bibitem{blaschke2016neutrino}
Blaschke DN, Cirigliano V.
\newblock \textit{Phys. Rev. D} 94(3):033009 (2016)

\bibitem{richers2019neutrino}
Richers SA, McLaughlin GC, Kneller JP, Vlasenko A.
\newblock \textit{Phys. Rev. D} 99(12):123014 (2019)

\bibitem{friedland2003neutrino}
Friedland A, Lunardini C.
\newblock \textit{Phys. Rev. D} 68(1):013007 (2003)

\bibitem{bell2003speed}
Bell NF, Rawlinson AA, Sawyer R.
\newblock \textit{Phys. Lett. B} 573:86--93 (2003)

\bibitem{balantekin2006neutrino}
Balantekin A, Pehlivan Y.
\newblock \textit{J. Phys. G} 34(1):47 (2006)

\bibitem{martin2023equilibration}
Martin JD, Neill D, Roggero A, Duan H, Carlson J.
\newblock \textit{Phys. Rev. D} 108(12):123010 (2023)

\bibitem{bauer2023quantum}
Bauer CW, Davoudi Z, Balantekin AB, Bhattacharya T, Carena M, et~al.
\newblock \textit{PRX Quantum} 4(2):027001 (2023)

\bibitem{di2024quantum}
Di~Meglio A, Jansen K, Tavernelli I, Alexandrou C, Arunachalam S, et~al.
\newblock \textit{PRX Quantum} 5(3):037001 (2024)

\bibitem{xiong2022}
Xiong Z.
\newblock \textit{Phys. Rev. D} 105(10):103002 (2022)

\bibitem{roggero2022}
Roggero A, Rrapaj E, Xiong Z.
\newblock \textit{Phys. Rev. D} 106(4):043022 (2022)

\bibitem{martin2022}
Martin JD, Roggero A, Duan H, Carlson J, Cirigliano V.
\newblock \textit{Phys. Rev. D} 105(8):083020 (2022)

\bibitem{lacroix2022}
Lacroix D, Balantekin AB, Cervia MJ, Patwardhan AV, Siwach P.
\newblock \textit{Phys. Rev. D} 106(12):123006 (2022)

\bibitem{shalgar2023we}
Shalgar S, Tamborra I.
\newblock \textit{Phys. Rev. D} 107(12):123004 (2023)

\bibitem{johns2023neutrino}
Johns L.
\newblock \textit{Int. J. Mod. Phys. A} 39(30):2450122 (2024)

\bibitem{cirigliano2024neutrino}
Cirigliano V, Sen S, Yamauchi Y.
\newblock \textit{Phys. Rev. D} 110(12):123028 (2024)

\bibitem{kost2024once}
Kost A, Johns L, Duan H.
\newblock \textit{Phys. Rev. D} 109(10):103037 (2024)

\bibitem{dasgupta2008collective}
Dasgupta B, Dighe A.
\newblock \textit{Phys. Rev. D} 77(11):113002 (2008)

\bibitem{raffelt2007self}
Raffelt GG, Sigl G.
\newblock \textit{Phys. Rev. D} 75(8):083002 (2007)

\bibitem{johns2023thermodynamics}
Johns L.
\newblock \textit{arXiv preprint arXiv:2306.14982}  (2023)

\bibitem{fiorillo2024inhomogeneous}
Fiorillo DFG, Raffelt GG, Sigl G.
\newblock \textit{Phys. Rev. Lett.} 133(2):021002 (2024)

\bibitem{hannestad2006self}
Hannestad S, Raffelt GG, Sigl G, Wong YY.
\newblock \textit{Phys. Rev. D} 74(10):105010 (2006)

\bibitem{duan2007analysis}
Duan H, Fuller GM, Carlson J, Qian YZ.
\newblock \textit{Phys. Rev. D} 75(12):125005 (2007)

\bibitem{xiong2023symmetry}
Xiong Z, Wu MR, Qian YZ.
\newblock \textit{Phys. Rev. D} 108(4):043007 (2023)

\bibitem{fiorillo2023slow}
Fiorillo DFG, Raffelt GG.
\newblock \textit{Phys. Rev. D} 107(4):043024 (2023)

\bibitem{johns2020neutrino}
Johns L, Nagakura H, Fuller GM, Burrows A.
\newblock \textit{Phys. Rev. D} 101(4):043009 (2020)

\bibitem{padilla2022neutrino}
Padilla-Gay I, Tamborra I, Raffelt GG.
\newblock \textit{Phys. Rev. Lett.} 128(12):121102 (2022)

\bibitem{johns2023collisional}
Johns L, Rodriguez S.
\newblock \textit{arXiv preprint arXiv:2312.10340}  (2023)

\bibitem{dasgupta2008spectral}
Dasgupta B, Dighe A, Mirizzi A, Raffelt GG.
\newblock \textit{Phys. Rev. D} 77:113007 (2008)

\bibitem{fogli2007collective}
Fogli G, Lisi E, Marrone A, Mirizzi A.
\newblock \textit{J. Cosmol. Astropart. Phys.} 2007(12):010 (2007)

\bibitem{raffelt2007self2}
Raffelt GG, Smirnov AY.
\newblock \textit{Phys. Rev. D} 76(8):081301 (2007)

\bibitem{raffelt2007adiabaticity}
Raffelt GG, Smirnov AY.
\newblock \textit{Phys. Rev. D} 76(12):125008 (2007)

\bibitem{dasgupta2009multiple}
Dasgupta B, Dighe A, Raffelt GG, Smirnov AY.
\newblock \textit{Phys. Rev. Lett.} 103(5):051105 (2009)

\bibitem{pehlivan2017spectral}
Pehlivan Y, Suba{\c{s}}{\i} A, Ghazanfari N, Birol S, Y{\"u}ksel H.
\newblock \textit{Phys. Rev. D} 95(6):063022 (2017)

\bibitem{wu2015physics}
Wu MR, Duan H, Qian YZ.
\newblock \textit{Phys. Lett. B} 752:89--94 (2016)

\bibitem{malkus2015symmetric}
Malkus A, McLaughlin GC, Surman R.
\newblock \textit{Phys. Rev. D} 93(4):045021 (2016)

\bibitem{zhu2016matter}
Zhu YL, Perego A, McLaughlin GC.
\newblock \textit{Phys. Rev. D} 94(10):105006 (2016)

\bibitem{frensel2016neutrino}
Frensel M, Wu MR, Volpe C, Perego A.
\newblock \textit{Phys. Rev. D} 95(2):023011 (2017)

\bibitem{shalgar2017multi}
Shalgar S.
\newblock \textit{J. Cosmol. Astropart. Phys.} 02:010 (2018)

\bibitem{padilla-gay:2024symmetry}
Padilla-Gay I, Shalgar S, Tamborra I.
\newblock \textit{J. Cosmol. Astropart. Phys.} 05:037 (2024)

\bibitem{capozzi2017fast}
Capozzi F, Dasgupta B, Lisi E, Marrone A, Mirizzi A.
\newblock \textit{Phys. Rev. D} 96(4):043016 (2017)

\bibitem{capozzi2019fast}
Capozzi F, Raffelt G, Stirner T.
\newblock \textit{J. Cosmol. Astropart. Phys.} 2019(09):002 (2019)

\bibitem{yi2019dispersion}
Yi C, Ma L, Martin JD, Duan H.
\newblock \textit{Phys. Rev. D} 99(6):063005 (2019)

\bibitem{fiorillo2024theory}
Fiorillo DFG, Raffelt GG.
\newblock \textit{J. High Energ. Phys.} 2024(225) (2024)

\bibitem{fiorillo2024theoryfast}
Fiorillo DFG, Raffelt GG.
\newblock \textit{J. High Energ. Phys.} 2024(12):1--28 (2024)

\bibitem{fiorillo2025collective}
Fiorillo DFG, Raffelt GG.
\newblock \textit{arXiv preprint arXiv:2502.06935}  (2025)

\bibitem{morinaga2022fast}
Morinaga T.
\newblock \textit{Phys. Rev. D} 105(10):L101301 (2022)

\bibitem{dasgupta2022collective}
Dasgupta B.
\newblock \textit{Phys. Rev. Lett.} 128(8):081102 (2022)

\bibitem{johns2024ergodicity}
Johns L.
\newblock \textit{arXiv preprint arXiv:2402.08896}  (2024)

\bibitem{raffelt2008self}
Raffelt GG.
\newblock \textit{Phys. Rev. D} 78(12):125015 (2008)

\bibitem{wolfenstein1979neutrino}
Wolfenstein L.
\newblock \textit{Phys. Rev. D} 20(10):2634 (1979)

\bibitem{esteban2008role}
Esteban-Pretel A, Mirizzi A, Pastor S, Tom{\`a}s R, Raffelt GG, et~al.
\newblock \textit{Phys. Rev. D} 78(8):085012 (2008)

\bibitem{chakraborty2011no}
Chakraborty S, Fischer T, Mirizzi A, Saviano N, Tomas R.
\newblock \textit{Phys. Rev. Lett.} 107(15):151101 (2011)

\bibitem{sigl2022}
Sigl G.
\newblock \textit{Phys. Rev. D} 105(4):043005 (2022)

\bibitem{shalgar2025neutrino}
Shalgar S, Tamborra I.
\newblock \textit{arXiv preprint arXiv:2503.03835}  (2025)

\bibitem{padilla2021fast}
Padilla-Gay I, Shalgar S.
\newblock \textit{arXiv preprint arXiv:2108.00012}  (2021)

\bibitem{sawyer2005speed}
Sawyer RF.
\newblock \textit{Phys. Rev. D} 72(4):045003 (2005)

\bibitem{mangano2014damping}
Mangano G, Mirizzi A, Saviano N.
\newblock \textit{Phys. Rev. D} 89(7):073017 (2014)

\bibitem{johns2020fast}
Johns L, Nagakura H, Fuller GM, Burrows A.
\newblock \textit{Phys. Rev. D} 102(10):103017 (2020)

\bibitem{bhattacharyya2020late}
Bhattacharyya S, Dasgupta B.
\newblock \textit{Phys. Rev. D} 102(6):063018 (2020)

\bibitem{bhattacharyya2020fast}
Bhattacharyya S, Dasgupta B.
\newblock \textit{Phys. Rev. Lett.} 126(6):061302 (2021)

\bibitem{wu2021collective}
Wu MR, George M, Lin CY, Xiong Z.
\newblock \textit{Phys. Rev. D} 104(10):103003 (2021)

\bibitem{duan2021flavor}
Duan H, Martin JD, Omanakuttan S.
\newblock \textit{Phys. Rev. D} 104(12):123026 (2021)

\bibitem{richers2021neutrino}
Richers S, Willcox D, Ford N.
\newblock \textit{Phys. Rev. D} 104(10):103023 (2021)

\bibitem{richers2022code}
Richers S, Duan H, Wu MR, Bhattacharyya S, Zaizen M, et~al.
\newblock \textit{Phys. Rev. D} 106(4):043011 (2022)

\bibitem{urquilla2024chaos}
Urquilla E, Richers S.
\newblock \textit{Phys. Rev. D} 109(10):103040 (2024)

\bibitem{xiong2023evaluating}
Xiong Z, Wu MR, Abbar S, Bhattacharyya S, George M, Lin CY.
\newblock \textit{Phys. Rev. D} 108(6):063003 (2023)

\bibitem{zaizen2022simple}
Zaizen M, Nagakura H.
\newblock \textit{Phys. Rev. D} 107(10):103022 (2023)

\bibitem{richers2024asymptoic}
Richers S, Froustey J, Ghosh S, Foucart F, Gomez J.
\newblock \textit{Phys. Rev. D} 110(10):103019 (2024)

\bibitem{george2024evolution}
George M, Xiong Z, Wu MR, Lin CY.
\newblock \textit{Phys. Rev. D} 110(12):123018 (2024)

\bibitem{abbar2023physics}
Abbar S, Wu MR, Xiong Z.
\newblock \textit{Phys. Rev. D} 109(4):043024 (2024)

\bibitem{abbar2024application}
Abbar S, Wu MR, Xiong Z.
\newblock \textit{Phys. Rev. D} 109(8):083019 (2024)

\bibitem{zaizen2023characterizing}
Zaizen M, Nagakura H.
\newblock \textit{Phys. Rev. D} 107(12):123021 (2023)

\bibitem{zaizen2024fast}
Zaizen M, Nagakura H.
\newblock \textit{Phys. Rev. D} 109(8):083031 (2024)

\bibitem{lin2023collision}
Lin YC, Duan H.
\newblock \textit{Phys. Rev. D} 107(8):083034 (2023)

\bibitem{kato2023collisional}
Kato C, Nagakura H, Johns L.
\newblock \textit{Phys. Rev. D} 109(10):103009 (2024)

\bibitem{zaizen2025spectral}
Zaizen M.
\newblock \textit{arXiv preprint arXiv:2502.09260}  (2025)

\bibitem{johns2022collisional}
Johns L, Xiong Z.
\newblock \textit{Phys. Rev. D} 106(10):103029 (2022)

\bibitem{johns2024subgrid}
Johns L.
\newblock \textit{arXiv preprint arXiv:2401.15247}  (2024)

\bibitem{nagakura2022general}
Nagakura H.
\newblock \textit{Phys. Rev. D} 106(6):063011 (2022)

\bibitem{nagakura2022time}
Nagakura H, Zaizen M.
\newblock \textit{Phys. Rev. Lett.} 129(26):261101 (2022)

\bibitem{shalgar2022neutrino}
Shalgar S, Tamborra I.
\newblock \textit{Phys. Rev. D} 108(4):043006 (2023)

\bibitem{shalgar2023neutrino}
Shalgar S, Tamborra I.
\newblock \textit{Phys. Rev. D} 107(6):063025 (2023)

\bibitem{nagakura2025resolution}
Nagakura H, Zaizen M, Liu J, Johns L.
\newblock \textit{Phys. Rev. D} 111(4):043028 (2025)

\bibitem{shalgar2019neutrino}
Shalgar S, Padilla-Gay I, Tamborra I.
\newblock \textit{J. Cosmol. Astropart. Phys.} 06:048 (2020)

\bibitem{padillagay2021multi}
Padilla-Gay I, Shalgar S, Tamborra I.
\newblock \textit{J. Cosmol. Astropart. Phys.} 2021(01):017 (2021)

\bibitem{nagakura2023bhatnagar}
Nagakura H, Johns L, Zaizen M.
\newblock \textit{Phys. Rev. D} 109(8):083013 (2024)

\bibitem{xiong2024fast}
Xiong Z, Wu MR, George M, Lin CY, Largani NK, et~al.
\newblock \textit{Phys. Rev. D} 109(12):123008 (2024)

\bibitem{nagakura2022connecting}
Nagakura H, Zaizen M.
\newblock \textit{Phys. Rev. D} 107(6):063033 (2023)

\bibitem{nagakura2023global}
Nagakura H.
\newblock \textit{Phys. Rev. D} 108(10):103014 (2023)

\bibitem{xiong2024robust}
Xiong Z, Wu MR, George M, Lin CY.
\newblock \textit{Phys. Rev. Lett.} 134(5):051003 (2025)

\bibitem{strack2005generalized}
Strack P, Burrows A.
\newblock \textit{Phys. Rev. D} 71(9):093004 (2005)

\bibitem{duan2008simulating}
Duan H, Fuller GM, Carlson J.
\newblock \textit{Comput. Sci. Discov.} 1(1):015007 (2008)

\bibitem{sarikas2012spurious}
Sarikas S, de~Sousa~Seixas D, Raffelt G.
\newblock \textit{Phys. Rev. D} 86(12):125020 (2012)

\bibitem{morinaga2018linear}
Morinaga T, Yamada S.
\newblock \textit{Phys. Rev. D} 97(2):023024 (2018)

\bibitem{duan2014multipole}
Duan H, Shalgar S.
\newblock \textit{J. Cosmol. Astropart. Phys.} 2014(10):084 (2014)

\bibitem{myers2022neutrino}
Myers M, Cooper T, Warren M, Kneller J, McLaughlin G, et~al.
\newblock \textit{Phys. Rev. D} 105(12):123036 (2022)

\bibitem{grohs2023neutrino}
Grohs E, Richers S, Couch SM, Foucart F, Kneller JP, McLaughlin G.
\newblock \textit{Phys. Lett. B} 846:138210 (2023)

\bibitem{grohs2023twomoment}
Grohs E, Richers S, Couch SM, Foucart F, Froustey J, et~al.
\newblock \textit{Astrophys. J.} 963(1):11 (2024)

\bibitem{froustey2023neutrino}
Froustey J, Richers S, Grohs E, Flynn SD, Foucart F, et~al.
\newblock \textit{Phys. Rev. D} 109(4):043046 (2024)

\bibitem{kneller2024quantum}
Kneller JP, Froustey J, Grohs EB, Foucart F, McLaughlin GC, Richers S.
\newblock \textit{arXiv preprint arXiv:2410.00719}  (2024)

\bibitem{froustey2024quantum}
Froustey J, Kneller JP, McLaughlin GC.
\newblock \textit{arXiv preprint arXiv:2409.05807}  (2024)

\bibitem{grohs2025advection}
Grohs E, Richers S, Froustey J, Foucart F, Kneller JP, McLaughlin GC.
\newblock \textit{arXiv preprint arXiv:2501.07540}  (2025)

\bibitem{fiorillo2024fast}
Fiorillo DFG, Raffelt GG.
\newblock \textit{Phys. Rev. Lett.} 133(22):221004 (2024)

\bibitem{liu2024quasi}
Liu J, Nagakura H, Zaizen M, Johns L, Akaho R, Yamada S.
\newblock \textit{Phys. Rev. D} 111(2):023051 (2025)

\bibitem{cornelius2023perturbing}
Cornelius M, Shalgar S, Tamborra I.
\newblock \textit{J. Cosmol. Astropart. Phys.} 02:038 (2024)

\end{thebibliography}

\end{document}